\documentclass[a4paper,12pt]{article}            
\usepackage{jheppub}
\pdfoutput=1

\usepackage[utf8]{inputenc}
\usepackage{amsmath,amssymb}
\usepackage{xcolor}
\usepackage{tensor}
\usepackage{graphicx}
\usepackage{subcaption}
\usepackage[normalem]{ulem}
\usepackage{xurl} 
\usepackage{hyperref}
\usepackage{bbm}
\usepackage{cancel}
\usepackage{mathtools}

\newcommand*\dif{\mathop{}\!\mathrm{d}} 
\newcommand{\xint}{\int \dif^{d+1}x \;}
\newcommand{\pint}{\int \frac{\dif^{d+1}p}{(2\pi)^{d+1}} \;}

\newcommand{\Jcal}{\mathcal{J}}
\newcommand{\texp}[1]{\;\text{exp}\left[#1\right]}
\newcommand{\veps}{\varepsilon}

\title{Schwinger-Keldysh effective actions for non-hydrodynamic poles and branch cuts}
\author[1,2]{Andrea Amoretti,}
\author[1,2]{Matteo Anselmi,} 
\author[1,2]{Daniel K. Brattan}

\emailAdd{andrea.amoretti@ge.infn.it}
\emailAdd{matteo.anselmi@ge.infn.it}
\emailAdd{danny.brattan@gmail.com}

\affiliation[1]{Dipartimento di Fisica, Universit\`a di Genova,
via Dodecaneso 33, I-16146, Genova, Italy}
\affiliation[2]{I.N.F.N. - Sezione di Genova, via Dodecaneso 33, I-16146, Genova, Italy}

\begin{abstract}
{\ Retarded thermal correlators, analytically continued to complex frequency, exhibit non-hydrodynamic poles and branch cuts that constrain hydrodynamics. We construct Gaussian Schwinger-Keldysh effective actions in which branch cuts arise from continua of relaxation poles and response functions are governed by Stieltjes transforms of relaxation-time densities. Dynamical KMS symmetry, positivity and stability constrain the spectra, while endpoint behaviour controls the local branch structure and near-edge pole scaling. The illustrative examples of uniform and power-law densities exhibit respectively beyond-all-orders pole-endpoint separation and a continuous interpolation between pole-pole and pole-branch-point termination. Finally, we show that an energy-dependent relaxation-time approximation in kinetic theory provides a microscopic realization of our branch cuts at zero momentum.}
\end{abstract}

\begin{document}

\maketitle

\section{Introduction}

Hydrodynamics is the universal effective description of the late-time, long-wavelength relaxation of conserved densities in a thermal medium~\cite{Kovtun:2012rj}. Its modern formulation is as an effective field theory (EFT) defined on the Schwinger-Keldysh (SK) closed time path~\cite{Crossley:2015evo,Glorioso:2017fpd,Haehl:2015pja,Haehl:2018lcu,Liu:2018kfw,Jain:2023obu}, where the doubling of fields incorporates dissipation and statistical fluctuations within the same framework. Two structural requirements play a central role. Dynamical KMS symmetry implements the thermal properties of the underlying state and enforces fluctuation-dissipation relations among the couplings, while positivity of the imaginary part of the action, inherited from microscopic unitarity, constrains the noise sector~\cite{Amoretti:2026positivenoise}. Retarded correlators of conserved densities are then organized around hydrodynamic poles whose dispersion relations can be computed order by order in gradients.

The physical retarded response of a system at real frequency is the boundary value $G^R(\omega+i0^+,\mathbf{k})$ of a function analytic in the upper half frequency plane. Its poles and branch points are singularities of the analytic continuation of this same causal response to generic complex frequencies. With the convention $e^{-i\omega t}$, isolated poles correspond to discrete damped contributions at $t>0$, while a branch cut contributes an integral over its discontinuity. In particular, hydrodynamic poles are characterized by approaching the origin of the complex frequency plane as the spatial momentum tends to zero. However, thermal correlators generally contain much more physics than can be described by just the hydrodynamic poles.

For example, at strong coupling, holographic theories exhibit towers of gapped quasinormal modes. As the coupling is reduced, these poles can also migrate and reorganize into branch cuts~\cite{Grozdanov:2016vgg}, which are characteristic features of weakly coupled kinetic theory~\cite{Romatschke:2015gic,Kurkela:2017xis,Moore:2018mma}. Sometimes one encounters weakly gapped ``quasihydrodynamic'' modes - i.e. long-lived non-hydrodynamic excitations - which, in addition to hydrodynamic poles, are important for describing late time physics. Such modes arise whenever a symmetry is softly broken or an approximately conserved quantity relaxes slowly~\cite{Grozdanov:2018fgx,Baggioli:2019jcm}, as in pinned collective modes and relaxing phonons in density-wave phases~\cite{Amoretti:2017xto,Amoretti:2018tzw,Amoretti:2019cef,Amoretti:2021lll}. The Maxwell-Cattaneo (MC) model of relaxing charge and heat transport~\cite{Cattaneo1948,Muller:1967zza,Israel:1979wp} provides the simplest representative of this class. More generally, anomalous power-law response in strongly correlated systems naturally points towards analytic structures that cannot be represented by a finite set of isolated poles.

The non-hydrodynamic sector also controls the regime of validity of hydrodynamics. Hydrodynamic dispersion relations $\omega(k)$ are analytic near $k=0$ but generally have a finite radius of convergence determined by non-analytic points of the spectral curve~\cite{Withers:2018srf,Grozdanov:2019kge,Grozdanov:2019uhi,Heller:2020hnq,Heller:2020uuy}. For discrete spectra, a familiar obstruction is the collision of a hydrodynamic pole with a gapped excitation. Branch cuts additionally allow direct encounters with branch points and collisions involving modes reached by analytic continuation to other sheets~\cite{Heller:2020hnq}.

Although the analytic structure of these continuum excitations has been studied extensively in kinetic theory~\cite{Bajec:2024rta,Brants:2024kin,Ochsenfeld:2023fjm,Hu:2024rta}, a systematic Schwinger-Keldysh description of non-hydrodynamic continua and their statistical sector remains less developed. Recent constructions reproduce propagating kinetic continua~\cite{Abbasi:2024rti,An:2025shift}, while complementary approaches work directly with correlators~\cite{Amoretti:2025kem} or derive quasihydrodynamic equations without the full statistical sector~\cite{Gavassino:2026ykg}. We construct such non-hydrodynamic structures directly within a Schwinger-Keldysh effective theory. We first generalize the  Maxwell-Cattaneo construction of~\cite{Jain:2023obu} to a conserved $U(1)$ charge coupled to $N$ relaxing vector fields. After imposing dynamical KMS symmetry, positivity and a frame choice for the current, the longitudinal spectrum contains diffusion together with $N$ non-hydrodynamic poles, and the conductivity is a sum of Drude-like terms weighted by their overlap with the physical current. 

Promoting the discrete relaxation spectrum to a continuum labelled by a relaxation time $\tau$ produces, for a natural class of couplings, a continuous weighted superposition of relaxation responses. This relaxation spectrum and its current overlaps are thereby promoted to EFT data, together with the fluctuation sector supplied by the SK construction. Retarded analyticity fixes the physical sheet, while the particular paths chosen as branch cuts are conventional. KMS symmetry, positivity and stability constrain the zero-momentum relaxation spectrum and allow paired off-axis branch points when the discrete symmetry permits antisymmetric structures. The behaviour of the relaxation-time density near its upper edge $\tau_*$ determines the local branch singularity and the scaling of nearby poles.
Their global trajectories, however, depend on the complete kernel.

We illustrate these statements in two analytically tractable families. A uniform relaxation-time density produces a logarithmic branch point and an isolated pole whose perturbative dispersion agrees with that of the continuum endpoint, while the two remain separated by an exponentially small contribution beyond all orders in $k^2$. At larger real momentum this pole collides with diffusion, producing a square-root singularity that bounds the convergence radius of the hydrodynamic series from above. On the other hand, in an exact power-law family, the complete spectral curve can be solved parametrically and the pole-pole collision approaches the continuum endpoint continuously as the endpoint exponent reaches $m=2$. Smooth local  gradient corrections move the endpoint while preserving its local universality class, whereas more general mixing and propagation introduce additional finite-momentum information. Finally, we provide a microscopic realization of the zero-momentum Stieltjes kernel using an energy-dependent relaxation time model in kinetic theory. 

The paper is organised as follows: Section~\ref{Sec:: Multiple non hydro poles} constructs the SK theory for finitely many relaxing modes and derives its spectrum, conductivity and heavy-mode reduction. Section~\ref{Sec:: Single branch cut} takes the continuum limit and studies its analytic structure through the uniform and power-law examples, while Sections~\ref{Sec:: Multiple branch cuts} and \ref{Sec:: Microscopic realization} address multiple cuts and a microscopic relaxation-time realization. We conclude in Section~\ref{Sec:: Conclusions}, and Appendix~\ref{App: Integration of heavy modes} gives explicit heavy-mode integrations.

\section{Multiple non-hydrodynamic poles}\label{Sec:: Multiple non hydro poles}
Following the Maxwell-Cattaneo (MC) model~\cite{Cattaneo1948,Muller:1967zza,Israel:1979wp,Grozdanov:2018fgx,Jain:2023obu,Amoretti:2026nonlin}, we augment SK diffusion for a conserved $U(1)$ current $J^\mu$ by $N$ additional vector fields $v_n^\mu$ each of which provides a local description of a non-hydrodynamic relaxation mode. The charge current couples to the usual gauge-invariant combination
  \begin{equation}
  B_\mu=A_\mu+\partial_\mu\varphi
\end{equation} 
with $A_\mu$ an external $U(1)$ gauge field, $\varphi$ a Stueckelberg field, and the $v_n$ are unsourced and satisfy $v_r^\mu u_\mu=0$. In the equilibrium frame we write $\mu_r=u^\mu B_{r\mu}$, with $\delta n=\chi\,\delta\mu$, where $\chi$ is the static susceptibility. We use the standard $r/a$ doubling and work at Gaussian order about homogeneous thermal equilibrium i.e. the action is quadratic in fluctuations, deterministic dynamics are linear, and statistical fluctuations are Gaussian. At this order the $ar$ terms encode linear response and the $aa$ terms are the KMS-related Gaussian noise.

To construct the action, in addition to imposing gauge invariance, we require the action to have a dynamical $\text{KMS}_\Theta$ symmetry in the classical limit
\begin{equation}
    \Phi_r \longrightarrow \Theta \Phi_r \;, \qquad \Phi_a \longrightarrow \Theta \Phi_a +i\Theta \beta^\mu\partial_\mu \Phi_r\;,
\end{equation}
where $\beta^\mu = \frac{u^\mu}{T}$ is the thermal vector, $\Theta$ is a discrete symmetry that contains time reversal and $\Phi$ is a field of the model. We also require invariance under the time-independent diagonal shift symmetry
\begin{equation}
    B_{r,\mu}\longrightarrow B_{r,\mu} +\partial_\mu \lambda(\vec{x})\;, \qquad B_{a\mu} \longrightarrow B_{a\mu}\;,
\end{equation}
that puts us in the unbroken phase of the $U(1)$ symmetry\footnote{A complete discussion on the emergence and the (non)necessity of this symmetry is shown in~\cite{Firat:2025upx}.}.

With this said, building on the SK action for the MC model in~\cite{Jain:2023obu}, the most general action at leading order in spacetime derivatives, consistent with these symmetries, is:
\begin{equation}\label{eq:: MC N action}
    S_{MC}[B_{s\mu},v_{n,s}^\mu] = \xint \Bigg\{ -\begin{pmatrix}
        B_{a\mu} & \vec{v}_{a\mu}^t
    \end{pmatrix} \mathbb{X}^{\mu\nu} \begin{pmatrix}
        B_{r\nu} \\
        \vec{v}_{r\nu}
    \end{pmatrix} + iT\begin{pmatrix}
        B_{a\mu} & \vec{v}_{a\mu}^t
    \end{pmatrix} \mathbb{S} \begin{pmatrix}
        \tilde{B}_{a\nu} \\
        \tilde{\vec{v}}_{a\nu}
    \end{pmatrix}\Bigg\} \;,
\end{equation}
where
    \begin{align}
        \mathbb{X}^{\mu\nu} =\begin{bmatrix}
        -\chi u^\mu u^\nu & 0 \\
        0 &X\Delta^{\mu\nu}
    \end{bmatrix} \qquad
    \mathbb{S} = \begin{bmatrix}
        -\lambda u^\mu u^\nu + \sigma \Delta^{\mu\nu} & (\vec{\gamma}_+^t + \vec{\gamma}_-^t)\Delta^{\mu\nu} \\
        (\vec{\gamma}_+ - \vec{\gamma}_-)\Delta^{\mu\nu} &\Sigma\Delta^{\mu\nu}
    \end{bmatrix}\;.
    \end{align}
Here, the tilde is the $\text{KMS}_\Theta$ transformed field and we have defined
\begin{equation}
    \vec{v}_{s\mu} = (v_{1,s\mu},\dots , v_{N,s\mu})^t\;, \qquad \Delta^{\mu\nu} = \eta^{\mu\nu}+u^\mu u^\nu\,.
\end{equation}
$\text{KMS}_\Theta$-invariance imposes conditions on the coefficient matrices:
\begin{subequations} \label{eq:: MC N KMS constraints of matrices}
    \begin{align}
        &X=\tilde{X}^t \\
        &\Sigma = \tilde{\Sigma}^t \\
        &\vec{\gamma}_\pm =\pm \tilde{\vec{\gamma}}_\pm\;.
    \end{align}
\end{subequations}
In addition, dynamical KMS invariance at this order requires
\begin{displaymath}
 \beta^\mu\partial_\mu X=0,
\end{displaymath}
which in the equilibrium rest frame reduces to $\partial_t X=0$. More general state- or spacetime-dependent coefficients require the corresponding nonlinear KMS completion, which is beyond the Gaussian action considered here. By looking at Table~\ref{Tab:: Transformation properties}, we see that the conditions depend on the inclusion of charge conjugation symmetry.

\begin{table}
\centering
\begin{tabular}{c|c|c|c|c|c|c}
    & C & P & T & CT & PT & CPT\\
    \hline
    $u^i$ & $+$ & $-$ & $-$ & $-$ & $+$ & $+$  \\
    $T$ & $+$ & $+$ & $+$ & $+$ & $+$ & $+$ \\ 
    $\mu$ & $-$ & $+$ & $+$ & $-$ & $+$ & $-$ \\ 
    $v^i$ & $-$ & $-$ & $-$ & $+$ & $+$ & $-$ \\ 
    $B_i$ & $-$ & $-$ & $-$ & $+$ & $+$ & $-$ 
\end{tabular}
\caption{Properties of transformation under discrete symmetries.}\label{Tab:: Transformation properties}
\end{table}

We identify a particular fixed combination of the relaxing fields in terms of the dissipative $U(1)$ current:
 \begin{equation}\label{eq:: MC N current frame choice}
    J^\mu = n u^\mu + \Jcal^\mu \;, \hspace{1cm} \Jcal^\mu = \sum_{n=1}^N \alpha_n v^\mu_n\equiv\sum_{n=1}^N \Jcal_n^\mu\;.
\end{equation}
Only $\vec\alpha^{\,t}\vec v^{\,\mu}$ will enter the physical current, leaving $\vec v^\mu\to R\vec v^\mu$ as a parametrization redundancy whenever $\vec\alpha^{\,t}R=\vec\alpha^{\,t}$. In a basis with $\vec\alpha^{\,t}=(1,0,\ldots,0)$ we find
 \begin{equation}
R=
\begin{pmatrix}
1 & 0\\
\vec b & A
\end{pmatrix},
\qquad A\in GL(N-1,\mathbb R),\qquad \vec b\in\mathbb R^{N-1} \; .
\end{equation}

The $\vec{v}$ are not just constrained by the identification \eqref{eq:: MC N current frame choice} but must satisfy their own deterministic equations obtained by varying the action with respect to $\vec{v}_{a\mu}^t$:
\begin{equation}
   -X \vec{v}^\mu_r + \Delta^{\mu\nu}\Sigma u^\rho \partial_\rho \vec{v}_{r\nu} + \Delta^{\mu\nu} (\vec{\gamma}_+-\vec{\gamma}_-) u^\rho \partial_\rho B_{r\nu} =0\;.  
\end{equation}
If we assume that  $\Sigma$ is invertible on the space it acts\footnote{When $\Sigma$ is only positive semidefinite, inverses are understood on its range or equivalently through the Moore-Penrose generalized inverse.} then we can isolate:
\begin{equation}
     \Delta^{\mu\nu} u^\rho \partial_\rho \vec{v}_{r\nu} = \Sigma^{-1}\Big(X \vec{v}^\mu_r - \Delta^{\mu\nu}(\vec{\gamma}_+ - \vec{\gamma}_-)u^\rho \partial_\rho B_{r\nu} \Big)\;.
\end{equation}
Using this equation only to eliminate $u^\rho\partial_\rho\vec v_{r\nu}$ -without integrating out the relaxing fields- the gauge current is:
\begin{align}
    J^\mu =&\; (n-\lambda u^\nu \partial_\nu \mu_r) u^\mu + \sigma \Delta^{\mu\nu} u^\rho \partial_\rho B_{r\nu} + \Delta^{\mu\nu} (\vec{\gamma}_+^t + \vec{\gamma}_-^t) u^\rho \partial_\rho \vec{v}_{r\nu} \nonumber \\
    =&\; (n-\lambda u^\nu \partial_\nu \mu_r) u^\mu + (\vec{\gamma}_+^t + \vec{\gamma}_-^t) \Sigma^{-1} X \vec{v}^\mu_r + \nonumber \\
    &+  \Big[ \sigma -(\vec{\gamma}_+^t + \vec{\gamma}_-^t) \Sigma^{-1}  (\vec{\gamma}_+ - \vec{\gamma}_-)\Big]\Delta^{\mu\nu} u^\rho \partial_\rho B_{r\nu}\;.
\end{align}
This imposes a consistency condition on the $\vec{v}$ and matching this current to \eqref{eq:: MC N current frame choice} requires that we impose:
\begin{subequations} \label{eq:: MC N constraints from frame choice}
\begin{align}
    \lambda &=0 \\
     (\vec{\gamma}_+^t + \vec{\gamma}_-^t) \Sigma^{-1} X&= \vec{\alpha}^t \label{eq:: MC N relabeling of alpha} \\
    (\vec{\gamma}_+^t + \vec{\gamma}_-^t) \Sigma^{-1}  (\vec{\gamma}_+ - \vec{\gamma}_-)&=\sigma\; . \label{eq:: MC N constraint on sigma}
\end{align}
\end{subequations}

Let us now count the number of degrees of freedom. Suppose from now on that we are not including charge conjugation in the discrete $\Theta$-symmetry, from~\eqref{eq:: MC N KMS constraints of matrices}, $\Sigma$ and $X$ must be symmetric and $\vec{\gamma}_-=0$, while $\vec{\gamma}_+$ is a generic vector in $\mathbbm{R}^N$. By including also $\chi, \sigma$ we have $N^2+2N+2$ free parameters. We now impose~\eqref{eq:: MC N constraints from frame choice}. In particular, the condition~\eqref{eq:: MC N relabeling of alpha} is just a relabelling that introduces $\vec{\alpha}^t$ and therefore does not reduce our degrees of freedom. The condition~\eqref{eq:: MC N constraint on sigma} however removes one independent parameter. Therefore, we have $(N+1)^2$ independent parameters, which is the exact number of parameters one would find by formulating the model starting from the entropy current~\cite{Jain:2023obu,Amoretti:2026nonlin}.

\subsection{Mode spectrum}
Let us now focus on the linearized equations of motion, obtained by varying the action~\eqref{eq:: MC N action} with respect to the $a$-fields. In Fourier space\footnote{We use the Fourier convention
\begin{equation}
    \phi(x) =\pint \phi(p) \texp{-i(\omega t-\vec{k}\cdot \vec{x})}\;.
\end{equation}
}:
\begin{subequations}\label{eq:: MC N Fourier EOMs}
    \begin{align}
        \omega \chi \delta\mu - k \vec{\alpha}^t \delta \vec{v}^{||} &=0 \\
        (X - i\omega \Sigma)\delta \vec{v}^{||} &= \vec{\gamma}_+ (\delta E^{||}-i k \delta \mu)\\
        (X - i\omega \Sigma) \delta \vec{v}^\perp &= \vec{\gamma}_+ \delta E^\perp \;,
    \end{align}
\end{subequations}
where we already separated the longitudinal and the transverse sectors. The transverse sector shows the emergence of the $N$ non-hydrodynamic modes:
\begin{equation}
    \delta \vec{v}^\perp = (X-i\omega\Sigma)^{-1} \vec{\gamma}_+ \delta E^\perp\;.
\end{equation}
This also enforces a physical meaning on the operator $M=X^{-1}\Sigma$ as a generalized relaxation time. If we now move to the longitudinal sector, we can easily solve for the chemical potential in the continuity equation, and by substituting it we obtain:
\begin{equation}
    \bar{M}\delta\vec{v}^{||} = \vec{\gamma}_+\delta E^{||}\;, \qquad \bar{M}=X-i\omega\Sigma+\frac{i k^2}{\omega\chi} \vec{\gamma}_+ \vec{\alpha}^t
\end{equation}
The longitudinal modes obey $\det\bar M=0$.  Therefore, as expected, in the longitudinal sector we have an additional mode. At finite momentum, these $N+1$ longitudinal modes satisfy:
\begin{equation}\label{Eq:Diffusive mode from matrix determinant lemma}
    \omega\det[X-i\omega\Sigma]+\frac{ik^2}{\chi} \vec{\alpha}^t \text{Adj}[X-i\omega\Sigma] \vec{\gamma}_+=0\;.
\end{equation}
We notice the presence of the diffusive mode, as the frequency admits a vanishing solution if the momentum is tuned to zero. 

Equation~\eqref{Eq:Diffusive mode from matrix determinant lemma} admits a suggestive rewriting: it is precisely the dispersion relation that follows from a Mori-Zwanzig memory equation for the charge density~\cite{Mori1965,Zwanzig1960},
\begin{equation} \label{Eq:Diffusive Mori Zwanzig}
    \chi\partial_t \delta \psi_D (t) = \nabla^2  \int \dif t' K(t-t') \delta \psi_D(t')\;,
\end{equation}
with memory kernel $K(\omega) =\vec{\alpha}^t (X-i\omega\Sigma)^{-1} \vec{\gamma}_+$. From this point of view, the relaxing vectors $v^\mu_n$ provide a local SK realization of the projected fast dynamics. The entire effect of the non-hydrodynamic sector on the conserved density is encoded in the analytic structure of $K(\omega)$. The continuum limit that we construct in Section~\ref{Sec:: Single branch cut}  then provides a natural realization of memory kernels with a continuous spectrum of relaxation times.

\paragraph{Effective theory for the diffusive mode}
The equation~\eqref{Eq:Diffusive mode from matrix determinant lemma} allows us to obtain an effective expression for the diffusive mode if we integrate the non-hydrodynamic modes out. Indeed:
\begin{equation}
    (X-i\omega\Sigma)^{-1} =(1-i\omega M)^{-1} X^{-1} \approx (1+i\omega M+\dots)X^{-1}
\end{equation}
where in the last step we have assumed that $\omega$ is small compared to all the typical scales given by $M^{-1}$. Therefore:
\begin{equation}
    \omega+\frac{ik^2}{\chi} \vec{\alpha}^t (X^{-1}+i\omega M X^{-1}+\dots)\vec{\gamma}_+=0\;.
\end{equation}
At leading order in $k$, we find the standard result:
\begin{equation}
    \omega_D(k) = -i\frac{\sigma}{\chi} k^2 + O(k^4)\;,
\end{equation}
where we have used the frame constraints~\eqref{eq:: MC N constraints from frame choice}. The first correction given by the heavy modes is:
\begin{equation}
    \omega_D(k) = -i\frac{\sigma}{\chi} k^2 \Big(1+c_{(4)} k^2\Big) + O(k^6)\;, \qquad c_{(4)}=\frac{1}{\chi}\,\vec{\alpha}^t M X^{-1} \vec{\gamma}_+ \;.
\end{equation}

\paragraph{Conductivity}
By looking at the equations~\eqref{eq:: MC N Fourier EOMs}, we can extract the conductivity
\begin{equation}\label{Eq:N modes conductivity resummed}
    \sigma(\omega) = \vec{\alpha}^t(1-i\omega M)^{-1} X^{-1}\vec{\gamma}_+\;.
\end{equation}
We can perform a spectral decomposition:
\begin{equation} \label{Eq:N modes conductivity}
    \sigma(\omega) = \sum_{a=1}^N \frac{\vec{\alpha}^t P_aX^{-1} \vec{\gamma}_+}{1-i\omega\tau_a} \;,
\end{equation}
where $\tau_a$ are the eigenvalues of $M$ and $P_a$ is the projector onto the corresponding eigenspace. We notice how the residue of a pole is governed by the product $\vec{\alpha}^t P_a$, i.e. by how much the eigenvector overlaps with the physical sector $\vec{\alpha}$. A mode with $\vec\alpha^{\,t}P_a=0$ is then absent from the conductivity. In particular, if $\vec\alpha^{\,t}$ is a left eigenvector of $M$, only the corresponding eigenspace contributes.

\subsection{Linear effective equations for the MC current}
We would like to express the linearized equations of motion obtained from the action~\eqref{eq:: MC N action} in terms of the physically relevant variables, that is, the chemical potential and the spatial current. Again, the linearized equations of motion in Fourier space read:
\begin{subequations}
    \begin{align}
        \omega \chi \delta\mu - k_i \vec{\alpha}^t \delta \vec{v}^i &=0 \\
        A(\omega)\delta \vec{v}^i &= \vec{B}\; (\delta E^i-i k^i \delta \mu) \;,
    \end{align}
\end{subequations}
where we have defined
\begin{equation}
    A(\omega) = \mathrm{1}-i\omega M\;, \qquad \vec{B} = X^{-1}\vec{\gamma}_+\;.
\end{equation}
Multiplying by the adjugate of $A(\omega)$ gives
\begin{equation}
    \det[A(\omega)] \delta\vec{v}^i = \text{Adj}[A(\omega)] \vec{B} \; (\delta E^i- ik^i \delta \mu) \;.
\end{equation}
We now notice that we can project along $\vec{\alpha}$ and obtain a closed equation for the physical current $\delta\Jcal^i = \vec{\alpha}^t\delta \vec{v}^i$. In particular, the adjugate of an $N$-dimensional matrix linear in $\omega$ can be expressed as a polynomial of degree $N-1$:
\begin{equation}
    \text{Adj}[A(\omega)] = C_0 + C_1\omega +\dots C_{N-1} \omega^{N-1}\;.
\end{equation}

Returning to position space, the physical variables must then obey
\begin{align}
    \chi \partial_t \delta\mu + \partial_i \delta \Jcal^i&=0 \; ,  \label{Eq:Continuity physical current}\\
    \prod_{a=1}^N \Big(1+\tau_a\partial_t\Big) \delta\Jcal^i &= \sum_{n=0}^{N-1} c_n \partial_t^n  (\delta E^i-\partial^i \delta \mu)\;. \label{Eq:Equation for physical current}
\end{align}
For $n\geq1$, charge conservation implies
\begin{equation}
    \partial_t^n\partial^i\delta\mu=-\frac{1}{\chi}\,\partial_t^{n-1}\partial^i\partial_j\delta\Jcal^j\; ,
\end{equation}
and thus we can eliminate higher time derivatives from the MC-like equation to find
\begin{align}\label{Eq:Closed physical current}
    \Bigg[&\prod_{a=1}^N(1+\tau_a\partial_t)\,\delta^i_{\ j}
    -\frac{1}{\chi}\sum_{n=1}^{N-1}c_n\partial_t^{n-1}\partial^i\partial_j\Bigg]\delta\Jcal^j + c_0\partial^i\delta\mu
    =\sum_{n=0}^{N-1}c_n\partial_t^n\delta E^i \;.
\end{align}
The highest time derivative acting on a dynamical field is of order $N$, so no extra relaxation poles have been introduced. Moreover we have isolated hydrodynamic variables on the left in \eqref{Eq:Closed physical current} and background couplings on the right. The derivatives of the external electric field (the background terms) generate the holomorphic numerator terms of the conductivity \cite{Amoretti:2025kem}. This holomorphic structure, however, is not simply a consequence of excluding heavy modes in a low-energy description, but can instead be given a physical interpretation in terms of the all-order external electric field. Such numerator data also become physically significant under duality transformations, which can interchange poles and zeros of linear-response functions~\cite{Amoretti:2026SL2Z}.

\subsection{Integration of heavy modes}
Finally, we note that parametrically fast non-hydrodynamic modes can be integrated out by a Gaussian Schur complement. The exact reduced action is generally non-local in spacetime, while a local EFT follows from a derivative expansion. Integrating out the complete relaxing sector yields the memory kernel~\eqref{Eq:Diffusive Mori Zwanzig} in the retarded sector, and the statistical kernel obeys the corresponding dynamical KMS relation. Appendix~\ref{App: Integration of heavy modes} gives explicit examples, including partial integrations that retain slower non-hydrodynamic modes.
 
\section{Single branch cut} \label{Sec:: Single branch cut}
We now extend the construction discussed above to a continuum of non-hydrodynamic poles by replacing the discrete vector label with a continuous parameter $\tau$. At Gaussian order the action becomes:

{\small
\begin{align}
    \label{Eq:Branchaction}
    S_{MC}[B_{s\mu},&v_{s\mu}(\tau)] = \xint \dif \tau \dif \tau'\Bigg\{ \begin{pmatrix}
    B_{a\mu} & v_{a\mu}(\tau)
    \end{pmatrix}
    \begin{bmatrix}
    \frac{\chi}{I^2} u^\mu u^\nu & 0 \\
    0 & -X(\tau,\tau')\, \Delta^{\mu\nu}
    \end{bmatrix}
    \begin{pmatrix}
    B_{r\nu} \\
    v_{r\nu}(\tau')
    \end{pmatrix} +\nonumber\\
    &+ iT
    \begin{pmatrix}
    B_{a\mu} &v_{a\mu}(\tau)
    \end{pmatrix}
    \begin{bmatrix}
    -\frac{\lambda}{I^2} u^\mu u^\nu + \frac{\sigma}{I^2} \Delta^{\mu\nu} & \frac{1}{I}(\gamma_+(\tau')+\gamma_-(\tau'))\, \Delta^{\mu\nu} \\
    \frac{1}{I}(\gamma_+(\tau)-\gamma_-(\tau))\, \Delta^{\mu\nu} & \Sigma(\tau,\tau')\, \Delta^{\mu\nu}
    \end{bmatrix}
    \begin{pmatrix}
    \tilde{B}_{a\nu} \\
    \tilde{v}_{a\nu}(\tau')
    \end{pmatrix}
    \Bigg\}\,.
\end{align}
}%
Here $I=\int d\tau$ is the length of the integration interval. For $0<\tau\leq\tau_\ast$, the continuum of
relaxation poles sit in the complex frequency plane at
\begin{equation}
\omega=-\frac{i}{\tau}.
\end{equation}
Thus $\tau\to0^+$ maps to $\omega\to-i\infty$, while the upper
endpoint $\tau=\tau_\ast$ maps to the finite branch point
\begin{equation}
\omega_\ast=-\frac{i}{\tau_\ast}.
\end{equation}

The $\text{KMS}_\Theta$ symmetry acting on our theory requires
\begin{equation}
    X(\tau,\tau')=\tilde{X}(\tau',\tau) \;, \quad \Sigma(\tau,\tau')=\tilde{\Sigma}(\tau',\tau)\;, \quad \gamma_\pm(\tau) = \pm\tilde{\gamma}_\pm(\tau)\;.
\end{equation}
Analogously to the discrete case, we impose a condition on the field $v^{\mu}(\tau)$ so that the gauge current has the form
\begin{equation}
    J^\mu = n u^\mu +\langle \alpha|v^\mu\rangle\; ,
\end{equation}
where $\langle \alpha|v^\mu\rangle =\int \dif\tau \; \alpha(\tau)v^\mu(\tau)$. Consistency with the equation of motion given by varying $v_{a}^{\mu}(\tau)$ then implies that
\begin{subequations} \label{eq:: Branch cut frame conditions}
\begin{align}
    \lambda &=0 \\
     \langle\gamma_+ + \gamma_-| \Sigma^{+} X&= \langle\alpha| \label{eq:: Branch cut gamma alpha frame constraint}\\
    \langle\gamma_+ + \gamma_-| \Sigma^{+}|\gamma_+ - \gamma_-\rangle&=\sigma\; ,
\end{align}
\end{subequations}
as we found for the discrete case. Here $\Sigma^{+}$ denotes the inverse on the dynamical range of $\Sigma$. For a strictly positive kernel it is the ordinary inverse. If zero lies in the continuous spectrum, $\Sigma^{+}$ is understood on its natural domain or, equivalently for the examples below, by first introducing a lower cutoff on the relaxation-time interval and removing it after the correlators have been formed.

\paragraph{Equations of motion}
The linearized equations of motion generated by this action \eqref{Eq:Branchaction} are:
\begin{align}
    \omega\chi\delta\mu - k_i\langle \alpha|\delta v^i\rangle&= 0 \; , \\
   \Big[X - i\omega\Sigma\Big]| \delta v^i\rangle &= \Big(\delta E^i - ik^i\delta\mu\Big)|\gamma_+\rangle\;.
\end{align}
The solution of these equations follows from the discrete case. Using our Fourier convention, in the equilibrium rest frame
$u\cdot\partial=\partial_t\rightarrow -i\omega$, if we define
\begin{equation}
    A(\omega;\tau,\tau')=X(\tau,\tau') - i\omega\Sigma(\tau,\tau')
\end{equation}
and use its resolvent away from its spectrum, the solutions are
\begin{subequations}
\label{Eq:Responses}
\begin{align}
    \delta\mu &= \frac{kB(\omega)}{\omega\chi + ik^2B(\omega)}\delta E_{||} \\
    \delta \Jcal^{||} &= \frac{\omega\chi B(\omega)}{\omega\chi + ik^2B(\omega)} \delta E_{||} \\
    \delta \Jcal^\perp &= B(\omega) \delta E^\perp \;,
\end{align}
\end{subequations}
where
\begin{equation}\label{eq:: Definition of B}
    B(\omega ) = \langle\alpha| A^{-1}(\omega) |\gamma_+\rangle\;.
\end{equation}

For real frequency, the response functions \eqref{Eq:Responses} are understood as retarded boundary values, $\omega\rightarrow\omega+i0^+$. Their continuation into complex frequency is therefore controlled by the analytic continuation of $B(\omega)$. We recognize that, at finite momentum, the modes are described by an equation analogous to~\eqref{Eq:Diffusive mode from matrix determinant lemma}:
\begin{equation}\label{Eq:Branch cut modes equation}
    \omega + \frac{ik^2}{\chi} B(\omega)=0\;.
\end{equation}
In the transverse channel $B(\omega)$ will directly determine the response, while in the longitudinal channel it enters both numerator and denominator. Any non-analyticities of $B(\omega)$ will consequently be inherited by the correlators, whereas isolated collective poles occur at zeros of the denominator, equivalently at solutions of \eqref{Eq:Branch cut modes equation}.

\subsection{Stieltjes representation and endpoint data}\label{Sec:: Stieltjies transform}
The mode equation~\eqref{Eq:Branch cut modes equation} is controlled by $B(\omega)$. We shall now specialise to a particular class of terms satisfying 
\begin{equation}\label{eq:: Stieltjies choice of X and Sigma}
    X(\tau,\tau')=\chi_v\delta(\tau-\tau')\;,\qquad
    \Sigma(\tau,\tau')=\chi_v\tau\delta(\tau-\tau')\;.
\end{equation}
This choice corresponds to making the kernels diagonal in the relaxation label. The delta functions in $\tau$ imply that modes with different relaxation times do not mix at this order, so that each value of $\tau$ labels an independent Maxwell-Cattaneo relaxation direction. However, if the support includes $\tau=0$, multiplication by $\tau$ is not invertible as a bounded operator. In such cases we define intermediate expressions on $I_{\tau_0}=[\tau_0,\tau_*]$ with $\tau_0>0$ and take $\tau_0\to0^+$ after computing finite-frequency observables. Since the endpoint of interest is $\tau_*$, this regulator does not affect the local branch structure discussed below. Equivalently, one may use the generalized inverse in~\eqref{eq:: Branch cut frame conditions} on its natural domain.

With no charge conjugation in $\Theta$, the frame constraints are solved by
\begin{equation}\label{eq:: Branch cut specific frame conditions}
    \gamma_+(\tau)=\alpha(\tau)\tau\;,\qquad
    \sigma=\frac{1}{\chi_v}\int_I\dif\tau\;\alpha^2(\tau)\tau\;.
\end{equation}
The second relation guarantees the Einstein relation for the diffusive mode. Defining
\begin{equation}
    f(\tau)\equiv\alpha(\tau)\gamma_+(\tau)=\alpha^2(\tau)\tau\;,
\end{equation}
we obtain
\begin{equation}\label{eq:: B Stieltjes representation}
    B(\omega)=\frac{i}{\omega\chi_v}S_f^I\!\left(-\frac{i}{\omega}\right)
    =\frac{1}{\chi_v}\int_I\dif\tau\;\frac{f(\tau)}{1-i\omega\tau}\;,
\end{equation}
where
\begin{equation}
    S_f^I(z)=\int_I\dif\tau\;\frac{f(\tau)}{\tau-z}\;,
    \qquad z\notin I
\end{equation}
is the Stieltjes transform of $f$. In this frame $f\geq0$, while $X$ and $\Sigma$ determine the spectrum of rates and $f$ measures its overlap with the physical current.

A generic endpoint is not characterized by its leading power alone. Let
\begin{equation}\label{eq:: General endpoint expansion}
    f(\tau)=(\tau_*-\tau)^m\Big[a_0+a_1(\tau_*-\tau)+a_2(\tau_*-\tau)^2+\cdots\Big],
    \qquad m>-1\;,
\end{equation}
with $a_0\neq0$ and the bracket analytic near $\tau_*$. After subtracting terms holomorphic at $z=\tau_*$, the transform has the local form
\begin{equation}\label{eq:: Stieltjies near branch point}
    S_f^I(z)=H(z)+
    \begin{cases}
        C_m\,(z-\tau_*)^m\big[a_0+O(z-\tau_*)\big]\;,
        &m\notin\mathbbm{Z}_{\geq0}\;,\\
        C_n\,(z-\tau_*)^n\log(z-\tau_*)\big[a_0+O(z-\tau_*)\big]\;,
        &m=n\in\mathbbm{Z}_{\geq0}\;,
    \end{cases}
\end{equation}
where $H$ is holomorphic and $C_m,C_n$ are non-zero constants whose phases depend on the branch convention. Every subleading coefficient $a_j$ generates a corresponding higher-power singular term. The same exponents appear in $B(\omega)$ because $z=-i/\omega$ is a local analytic change of variable near $\omega_*=-i/\tau_*$. The holomorphic remainder $H$ contains the complementary global information i.e. the finite value $B(\omega_*)$, when it exists, and the global pole trajectories depend on the full functional form of the density.

\subsection{Movable cuts and rigid branch points}\label{Sec:: Movable cuts}
The Stieltjes representation separates physical analytic data from a conventional choice of cut. The discontinuity across the original real-$\tau$ integration interval is
\begin{equation}
    S_f^I(x+i0^+)-S_f^I(x-i0^+)=2\pi i f(x)\;,
    \qquad x\in\mathring I\;.
\end{equation}
If $f$ admits an analytic continuation near the interior of $I$, the integration contour may be deformed in the complex $\tau$ plane while its endpoints are kept fixed. The resulting representation agrees with the original correlator on their common domain of analyticity, although the drawn cut moves to the image of the deformed contour. The cut placement is therefore conventional. The endpoints, any interior singularities of the continued density, the local discontinuity and the associated monodromy are invariant data.

In particular, the endpoint exponent $m$ in~\eqref{eq:: General endpoint expansion} is unchanged by a smooth contour deformation. Sheet labels should consequently be defined by homotopy classes of continuation paths from the retarded domain in the upper half frequency plane, rather than by a particular drawing of the cuts. The figures below use the undeformed real-$\tau$ contour only for convenience.

\subsection{Branch choices and the retarded germ}\label{Sec:: Spurious mode from wrong sheets}
A retarded correlator is specified by a germ analytic in the upper half frequency plane, together with its continuation. By the retarded germ we mean the locally defined analytic function fixed in the upper half frequency plane, from which the continuation is constructed. Near a logarithmic branch point one may write
\begin{equation}
    B(\omega)=H(\omega)+C(\omega)\log u(\omega)\;,
\end{equation}
where $H$, $C$ and $u$ are holomorphic and $u(\omega_*)=0$. Replacing $\log u$ by $\log u+2\pi i n$ represents the same continued function only if the holomorphic remainder is changed simultaneously, $H\to H-2\pi i n C$. Changing the logarithm while holding $H$ fixed instead defines a different germ.

The uniform density considered in the next subsection makes this explicit:
\begin{equation}\label{eq:: Uniform B for branch discussion}
    B_0(\omega)=\frac{i\bar f}{\omega\chi_v}\log(1-i\omega\tau_*)\;.
\end{equation}
It is finite at $\omega=0$. If one replaces the logarithm by $\log(1-i\omega\tau_*)+2\pi i n$ while keeping the prefactor unchanged, one obtains
\begin{equation}
    B_n(\omega)=B_0(\omega)-\frac{2\pi n\bar f}{\omega\chi_v}\;,
\end{equation}
which has a new pole at the origin and therefore is not the same low-frequency retarded germ. Inserting this expression into~\eqref{Eq:Branch cut modes equation} gives, with $c=\bar f/(\chi\chi_v)$,
\begin{equation}
    \omega_\pm=\pm k\sqrt{2\pi i n c}-\frac{i c\tau_*k^2}{2}+O(k^3)\;.
\end{equation}
For every $n\neq0$, one of these roots lies in the upper half plane. This instability diagnoses the changed or analytically continued expression, it is not a pole of the physical retarded boundary value and does not select the physical sheet. The physical sheet is fixed from the outset by continuation of $B_0$ from its retarded domain.

\subsection{Local emergence from an endpoint}\label{Sec:: Local endpoint emergence}
We now extract what follows from the endpoint germ alone. Let $q=k^2$, introduce the local coordinate $u=1-i\omega\tau_*$, and write the spectral equation as
\begin{equation}
    P(\omega,q)=\omega+\frac{iq}{\chi}B(\omega)=0\;.
\end{equation}
For $-1<m<0$, the singular term in~\eqref{eq:: Stieltjies near branch point} diverges and a near-edge solution, whenever its phase places it on the retarded sheet, obeys
\begin{equation}
    u\sim q^{1/|m|}=k^{2/|m|}\;.
\end{equation}
For $m=0$, the logarithmic divergence instead gives
\begin{equation}
    u\sim\exp\!\left[-\frac{A}{q}\right]
\end{equation}
with $A$ fixed by the leading discontinuity and by $\omega_*$. These are local statements: their coefficients and sheet assignment depend on the phase of the singular germ.

For $m>0$, $B$ has a finite endpoint value $B_*$. A root can meet the endpoint at real momentum only if
\begin{equation}\label{eq:: General endpoint threshold}
    q_*=k_*^2=\frac{i\chi\omega_*}{B_*}
\end{equation}
is real and positive. Expanding around $(u,q-q_*)=(0,0)$ gives an analytic term linear in $u$, a term proportional to $q-q_*$, and the non-analytic endpoint term. Hence
\begin{align}
    0<m<1:&\qquad u\sim(q-q_*)^{1/m}\;,\\
    m=1:&\qquad u\log u\sim q-q_*\;,\\
    m>1:&\qquad u\sim q-q_*\quad\text{generically}.
\end{align}
For $m>1$, the coefficient and the side of the threshold are controlled by the holomorphic remainder $H$. If its linear coefficient happens to vanish, subleading endpoint and global data decide the scaling. Thus the exponent by itself cannot determine whether a new pole is emitted, which sheet it occupies, or how the hydrodynamic branch eventually terminates.

\begin{table}[t]
    \centering
    \small
    \begin{tabular}{c|c|c|c|c}
        $m$ & Endpoint germ & $B_*$ & $q_*$ & Near-edge root \\
        \hline
        $(-1,0)$ & $u^m$ & divergent & $0$ & $u\sim q^{1/|m|}$ \\
        $0$ & $\log u$ & divergent & $0$ & $u\sim e^{-A/q}$ \\
        $(0,1)$ & $u^m$ & finite & \eqref{eq:: General endpoint threshold} & $u\sim(q-q_*)^{1/m}$ \\
        $1$ & $u\log u$ & finite & \eqref{eq:: General endpoint threshold} & $u\log u\sim q-q_*$ \\
        $(1,\infty)$ & subleading singularity & finite & \eqref{eq:: General endpoint threshold} & generically $u\sim q-q_*$ \\
    \end{tabular}
    \caption{Consequences of the local endpoint expansion. The table describes a root in a neighbourhood of the branch point, subject to the phase and positivity conditions stated in the text. Global continuation of these roots requires the full Stieltjes kernel.}
    \label{tab:: Summary Branch points}
\end{table}

We now turn to two indicative, fully specified densities for which the real-momentum spectral evolution can be determined. Whether the resulting real-momentum singularity is indeed the nearest singularity to $k=0$ in the complex momentum plane remains an open question.

\subsection{Example I: a uniform relaxation-time density}\label{Sec:: Log branch cut}
We first specify the density on the whole interval by setting $f(\tau)=\bar f$ for $0<\tau<\tau_*$. A convenient implementation is
\begin{equation}\label{eq:: Log branch cut alpha gamma choices}
    \alpha(\tau)=\sqrt{\frac{\bar f}{\tau}}\;,\qquad
    \gamma_+(\tau)=\sqrt{\bar f\tau}\;,
\end{equation}
together with~\eqref{eq:: Stieltjies choice of X and Sigma}; the lower-endpoint regulator described above is understood. The frame condition fixes $\bar f=\sigma\chi_v/\tau_*$. This complete choice determines both the logarithmic singularity and the holomorphic part of the kernel,
\begin{equation}
    B(\omega)=\frac{i\bar f}{\omega\chi_v}\log(1-i\omega\tau_*)\;,
\end{equation}
and the mode equation becomes
\begin{equation}\label{eq:: Log branch cut modes equation}
    \omega=\frac{c k^2}{\omega}\log(1-i\omega\tau_*)\;,
    \qquad c=\frac{\bar f}{\chi\chi_v}\;.
\end{equation}
Besides the diffusive root, the retarded continuation contains a pole exponentially close to the logarithmic endpoint. Its exact trajectory and later collision are properties of this fully specified uniform model.

The diffusive mode is simple to compute. At leading order in $k^2$, the ansatz $\omega_D(k)=-iDk^2+O(k^4)$ gives
\begin{equation}
    D=c\tau_*=\frac{\sigma}{\chi}\;,
\end{equation}
which is the Einstein relation. By contrast, an ordinary analytic ansatz for the edge mode, $\omega_b(k)=-\frac{i}{\tau_*}+iak^2+O(k^4)$, leads to
\begin{equation}
    -\frac{i}{\tau_*} +iak^2 = i ck^2\tau_* \log[ak^2 \tau_{*}]
\end{equation}
which cannot be satisfied by a $k$-independent coefficient $a$ as $k\to0$. The separation from the endpoint must therefore be non-perturbative in $k^2$.

\begin{figure}
    \centering
    \includegraphics[width=0.7\linewidth]{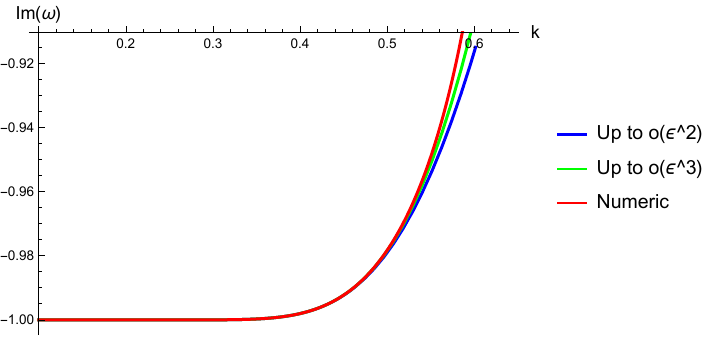}
    \caption{Numerical and analytic expressions for the non-perturbative pole emerging from the branch point. We set $c=\tau_*=1$.}
    \label{fig:: Plot Non-perturbative pole}
\end{figure}

For a solution with $\omega_b(k\to0)\to-i/\tau_*$, Equation~\eqref{eq:: Log branch cut modes equation} instead requires
\begin{equation}
    \log[1-i\omega_b(k)\tau_*]\sim-\frac{1}{c k^2\tau_*^2}\;.
\end{equation}
This motivates the ansatz
\begin{equation}
    \omega_b(k) = -\frac{i}{\tau_*} + \bar{\omega}_b(k) \texp{-\frac{\varkappa}{ck^2}}\;.
\end{equation}
From the $k\rightarrow0$ equation we obtain $\varkappa=1/\tau_*^2$ (we use a symbol distinct from the endpoint exponent $m$ introduced in Section~\ref{Sec:: Stieltjies transform}, with which it has nothing to do). By defining $\eta=1/ck^2\tau_*^2$, we obtain an equation for the amplitude $\bar{\omega}_b(k)$:
\begin{equation}\label{eq:: Equation for omega bar}
    \bar{\omega}^2_b(k) e^{-2\eta} - \frac{2i}{\tau_*} \bar{\omega}_b(k) e^{-\eta} = ck^2 \log[-i\bar{\omega}_b(k)\tau_*]\;.
\end{equation}
Define
\begin{equation}
    \veps = e^{-\eta} \;, \qquad \veps y=1-i\omega\tau_*\; ,
\end{equation}
so that \eqref{eq:: Log branch cut modes equation} becomes
\begin{equation}
    \log y = 2\eta \veps y -\eta \veps^2y^2\;.
\end{equation}
In this form, the equation allows for an asymptotic expansion of $y$. Writing
$y=y_0+y_1\veps+y_2\veps^2+O(\eta^3\veps^3)$, we obtain:
\begin{subequations}
    \begin{align}
        O(1):& \qquad y_0 =1 \\
        O(\veps):& \qquad y_1 = 2\eta \\
        O(\veps^2):& \qquad y_2 =6\eta^2 -\eta\;.
    \end{align}
\end{subequations}
Therefore, we get an expression for the non-perturbative pole:
\begin{equation}\label{Eq:Non perturbative mode}
    \omega_b(k) = -\frac{i}{\tau_*}
\Big[1-\veps-2\eta\veps^2-(6\eta^2-\eta)\veps^3
+O(\eta^3\veps^4)\Big] \; . \qquad
\end{equation}
Figure~\ref{fig:: Plot Non-perturbative pole} compares the numerical solution with this asymptotic expression.

Equation~\eqref{Eq:Non perturbative mode} admits a sharper reading once the branch point itself is allowed to move. In the strictly nondispersive model the endpoint of the continuum is fixed at $\omega_*=-i/\tau_*$, and the expansion above is non-perturbative in its entirety. Including the higher-gradient corrections of Section~\ref{Sec:: Gradient corrections}, the endpoint acquires an ordinary dispersion, $\omega_*(k)=-i/\tau_*(k)$ with $\tau_*(k)=\tau_*\left[1-k^2\ell^2(\tau_*)+O(k^4)\right]$, and the pole follows it,
\begin{equation}\label{eq:: Degenerate dispersions}
    \omega_b(k) = \omega_*(k) + O\!\left(e^{-1/(c(k)k^2\tau_*^2)}\right)\;, \qquad c(k) = c + \mathcal{O}(k^2) \; . \qquad
\end{equation}
Thus the endpoint and near-edge pole share all power-law terms in their dispersions and are distinguished only by an exponentially small splitting invisible to the gradient expansion, closely paralleling trans-series degeneracies \cite{Heller:2015dha,Aniceto:2018bis}. For $-1<m<0$ the corresponding separation is instead algebraic, $k^{2/|m|}$.

\subsubsection{Pole collision and the hydrodynamic branch}
For the uniform density, numerical continuation shows that the diffusive pole and the exponentially close edge pole merge. An analytic root branch can fail to remain analytic where the spectral equation itself is non-analytic or where a double root develops, $P=\partial_\omega P=0$~\cite{Heller:2020hnq}. The collision found here lies away from the logarithmic cut, so the double-root condition is sufficient and the generic local singularity is a square root.

\begin{figure}
    \centering
    \includegraphics[width=0.7\linewidth]{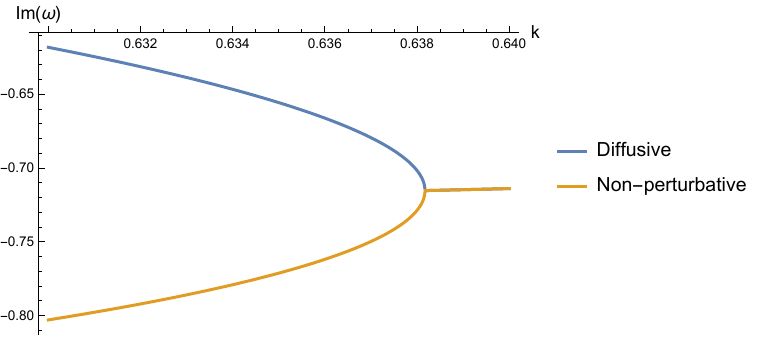}
    \caption{Imaginary parts of the two poles as functions of the momentum, in units $c=\tau_*=1$.}
    \label{fig:: Numeric collision}
\end{figure}

In our explicit case, 
\begin{equation}
    P(\omega,k) = \omega -\frac{ck^2}{\omega} \log[1-i\omega\tau_*]\;.
\end{equation}
We notice that this function is non-analytic only on the logarithmic branch cut, since the apparent singularity in $\omega=0$ can be eliminated with the logarithm. We change variables:
\begin{equation}
    u=1-i\omega\tau_*\;, \qquad q=\frac{1}{\eta} = ck^2\tau_*^2\;.
\end{equation}
The branch cut is then located along the negative real axis in the complex $z$ plane. Up to an overall constant factor:
\begin{equation}
    P(u,q) = 1-u+q\frac{\log u}{1-u}\;.
\end{equation}
For $u\neq1$,  the spectral curve is parametrized by
\begin{equation}
    q(u) = -\frac{(1-u)^2}{\log u}\;.
\end{equation}
On this curve, one can verify that
\begin{equation}
    \partial_u P(u,q)\Big|_{q=q(u)} = -2-\frac{1-u}{u\log u}\;,
\end{equation}
so that
\begin{equation}\label{eq:: u equation for IFT}
    \partial_u P(u,q)=0 \quad \Longleftrightarrow \quad 2u\log u = u-1\;.
\end{equation}
Equivalently, \eqref{eq:: u equation for IFT} is the condition $\dif q/\dif u=0$ on the spectral curve, i.e.\ the statement that two roots of the mode equation merge. For the real-momentum collision under consideration, $u\in(0,1)$.
Defining
\begin{equation}
    r=-\frac{1}{2u}<-\frac{1}{2},
    \label{eq:u_lambert}
\end{equation}
Equation~\eqref{eq:: u equation for IFT} becomes
\begin{equation}
    r e^r=-\frac{e^{-1/2}}{2}.
    \label{eq:lambert_equation}
\end{equation}
Its two real solutions are given by the Lambert branches $W_0$ and
$W_{-1}$. The principal branch gives
\begin{equation}
    W_0\left(-\frac{e^{-1/2}}{2}\right)=-\frac{1}{2},
\end{equation}
which corresponds to the excluded point $u=1$, while the $W_{-1}$
branch gives the non-trivial collision,
\begin{equation}
    r_c=W_{-1}\left(-\frac{e^{-1/2}}{2}\right).
    \label{eq:u_critical}
\end{equation}
Transforming back to the original variables therefore gives the critical point
\begin{equation}
    u_c=-\frac{1}{2r_c}\simeq 0.2846681,
    \qquad
    \omega_c\simeq-\frac{0.7153319\,i}{\tau_\ast},
    \qquad
    k_c=\frac{0.6381727}{\sqrt{c}\tau_\ast}.
\end{equation}
A direct numerical solution of~\eqref{eq:: Log branch cut modes equation}, shown in Figure~\ref{fig:: Numeric collision}, locates the merging of the two poles at $\sqrt{c}\tau_*k_c = 0.6381727$, in agreement with the closed-form result to all displayed digits.

As a final remark, we can determine the local analytic structure of the
collision. At the critical point
\begin{equation}
    P(u_c,q_c)=0\;, \qquad
    \partial_u P(u_c,q_c)=0\;,
\end{equation}
while $P_q$ and $P_{uu}$ are non-zero. Expanding around the critical point
therefore gives, at leading order,
\begin{equation}
    0=P_q\,\delta q+\frac{1}{2}P_{uu}\,\delta u^2
    +O(\delta q^{3/2})\;,
\end{equation}
where $\delta u=u-u_c$ and $\delta q=q-q_c$. Hence
\begin{equation}
    \delta u_\pm
    =\pm\sqrt{-\frac{2P_q}{P_{uu}}\,\delta q}
    +O(\delta q)\;.
\end{equation}
Since $q=ck^2\tau_*^2$ and $k_c\neq0$, one has
$\delta q\propto k-k_c$ near the critical point. The collision therefore
produces a square-root branch point of $\omega(k)$. Since this singularity occurs at $k=k_c$, the radius $R_k$ of the small-$k$ expansion satisfies $R_k\leq k_c$. Establishing equality would require excluding singularities at smaller $|k|$ elsewhere in the complex momentum plane.

The uniform-density model also provides a useful benchmark for the temporal
derivative expansion of the constitutive relation, within the EFT regime.
Expanding the memory kernel as
\begin{equation}
B(\omega)=\frac{\bar f\tau_*}{\chi_v}
\sum_{n\geq0}\frac{(i\omega\tau_*)^n}{n+1}\,,
\end{equation}
one finds a radius of convergence $|\omega|<1/\tau_*$, set by the logarithmic branch point. Since the pole-pole collision occurs at $|\omega_c|\tau_*\simeq0.7153$, it lies inside this domain and is recovered systematically by higher-order truncations of the constitutive relation. 
By contrast, the near-edge splitting probes the complementary, beyond-all-orders sector of the same analytic structure. Its separation from the continuum endpoint has an identically vanishing Taylor series about $k=0$, while $\omega_*$ lies on the boundary of the frequency expansion. Thus, the derivative expansion captures the finite-frequency pole-pole collision, but is blind to the exponentially small pole-edge separation, which is resolved only by the exact kernel.

\subsubsection{Residues}
Let us now consider the linear results~\eqref{Eq:Responses}; we compute the residue of the correlator:
\begin{equation}
    \langle J^0 J^0\rangle = \frac{ik^2 B(\omega)}{\omega + \frac{ik^2}{\chi}B(\omega)}\;.
\end{equation}
It is immediate to verify that both the isolated poles are simple for $k<k_{c}$. Thus, the residues at a simple pole $\omega_p$ can be expressed as
\begin{equation}
    \text{Res}_{\omega_p}[\langle J^0 J^0\rangle ] = -\frac{\chi \omega_p}{1+\frac{ik^2}{\chi}B'(\omega_p)}\;,
\end{equation}
where we have also used the defining property of the pole $ik^2B(\omega_p) = -\chi \omega_p$.

Explicitly, in our case
\begin{equation}
    B'(\omega) = \frac{\chi}{\omega}\Big[\frac{c\tau_*}{1-i\omega\tau_*} -\frac{ic}{\omega}\log[1-i\omega\tau_*]\Big]\;.
\end{equation}
When we compute it on the pole, we can use the fact that $\log[1-i\omega_p\tau_*] = \omega_p^2/ck^2$, which turns the logarithmic term of $B'$ into a pure constant, and obtain an exact expression for the residue:
\begin{equation}\label{eq:: Exact residue}
    1+\frac{ik^2}{\chi}B'(\omega_p) = 2+\frac{i\,ck^2\tau_*}{\omega_p z_p}
    \qquad\Longrightarrow\qquad
    \text{Res}_{\omega_p}[\langle J^0 J^0\rangle] = -\frac{\chi\,\omega_p^2 z_p}{2\,\omega_p z_p + i\,ck^2\tau_*}\;,
\end{equation}
where $z_p=1-i\omega_p\tau_*$. From this expression, we can compute perturbatively the residues:
\begin{subequations}
    \begin{align}
        \text{Res}_{\omega_D}[\langle J^0 J^0\rangle] &= i\chi c\tau_* k^2 + O(k^4) \\
        \text{Res}_{\omega_b}[\langle J^0 J^0\rangle]
&= -\frac{i\chi}{\tau_*}\eta\veps
+O(\eta^2\veps^2)\;.
    \end{align}
\end{subequations}
The first expression is the standard hydrodynamic result $\text{Res}_{\omega_D}=i\chi Dk^2$ with $D=c\tau_*$, as follows directly from~\eqref{Eq:Responses} at $B(\omega)\to\sigma$; the second is dominated by the $2\omega_pz_p$ term being negligible against $ick^2\tau_*$ near the branch point. At small momentum the contribution of the emergent pole in the correlator is therefore non-perturbatively small compared to the hydrodynamic one.

As the momentum increases, the near-edge pole evolves into an ordinary gapped
excitation. The two poles~\eqref{eq:: Exact residue} merge at $k_c$, where
$2\omega_p z_p+ick^2\tau_*=0$, and their residues diverge with opposite signs
as $(k_c-k)^{-1/2}$, as expected for the collision of two simple poles.
Numerically, in units $c=\tau_*=\chi=1$, the residue of the emergent pole
becomes comparable to that of the diffusive mode as the collision is
approached. At $k_c$, $\veps(k_c)\simeq0.0858$, showing that the pole has
already moved a finite distance away from the continuum endpoint. The
near-edge asymptotics~\eqref{Eq:Non perturbative mode} therefore describes its
small-momentum origin, while the exact solution follows its evolution into
the pole-pole collision.

\subsection{Example II: the power-law family}\label{Sec:: Algebraic branch cut}
The second complete example is the one-parameter family
\begin{equation}\label{eq:: Exact power law density}
    f(\tau)=\bar f(\tau_*-\tau)^m\;,
    \qquad -1<m<\infty\;,
\end{equation}
on the entire interval $[0,\tau_*]$. This pure-power choice fixes the full density, not only its endpoint exponent. With~\eqref{eq:: Stieltjies choice of X and Sigma}, the transform is
\begin{equation}
    B(\omega) =\frac{\bar{f}}{\chi_v} \frac{\tau_*^{1+m}}{1+m} \;\prescript{}{2}{F}_1(1,1;2+m;i\omega\tau_*)\;,
\end{equation}
where $\prescript{}{2}{F}_1(a,b;c;z)$ is the hypergeometric function. Its value at the branch point follows from the Gauss summation formula
\begin{equation}
    \prescript{}{2}{F}_1(a,b;c;1) =\frac{\Gamma(c)\Gamma(c-a-b)}{\Gamma(c-a) \Gamma(c-b)}\;, \qquad \Re[c-a-b]>0\;,
\end{equation}
where $\Gamma(z)$ is the Euler gamma function. Substituting our parameters, $B(\omega)$ diverges at the branch point for $m\leq0$, whereas for $m>0$
\begin{equation}
    B(\omega_*) = \frac{\bar{f}}{\chi_v} \frac{\tau_*^{1+m}}{m}\;.
\end{equation}
The critical value of the momentum becomes then:
\begin{equation}
    k_c =\sqrt{\frac{m}{c\tau_*^{2+m}}}\;,
\end{equation}
where we have again introduced $c=\bar{f}/\chi\chi_v$, that also carries the right dimension for $[c\tau_*^{2+m}] = [L]^{2}$ for every $m$.

We first verify explicitly how the local endpoint scalings are realized within this family. We consider the expansion:
\begin{equation}
    P\Big(\omega_*+\frac{i\veps}{\tau_*}, k_c + \delta k\Big) =\omega_*+ \frac{i\veps}{\tau_*} + ic(k_c+\delta k)^2 \frac{\tau_*^{1+m}}{1+m} \prescript{}{2}{F}_1(1,1;2+m;1-\veps)\;.
\end{equation}
Using the standard continuation formulas for the hypergeometric function~\cite{NIST:DLMF}, we obtain
\begin{equation}\label{eq:: 2F1 expansion}
    \prescript{}{2}{F}_1(1,1;2+m;1-\veps) \simeq \frac{m+1}{m} -\frac{(m+1)\pi}{\sin(m\pi)}\veps^m + \frac{m+1}{m(1-m)}\veps +O(\veps^{\min(2,m+1)})\;,
\end{equation}
which is valid for $m\notin \mathbb{Z}$. Therefore, we must distinguish two scenarios: $m\in(0,1)$ and $m>1$.

\begin{figure}
\centering
\begin{subfigure}{.5\textwidth}
  \centering
  \includegraphics[width=\linewidth]{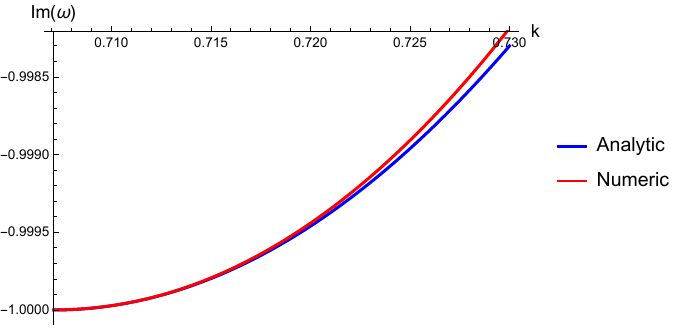}
  \caption{$m=\frac{1}{2}$}
  \label{subfig:: m=0.5}
\end{subfigure}%
\begin{subfigure}{.5\textwidth}
  \centering
  \includegraphics[width=\linewidth]{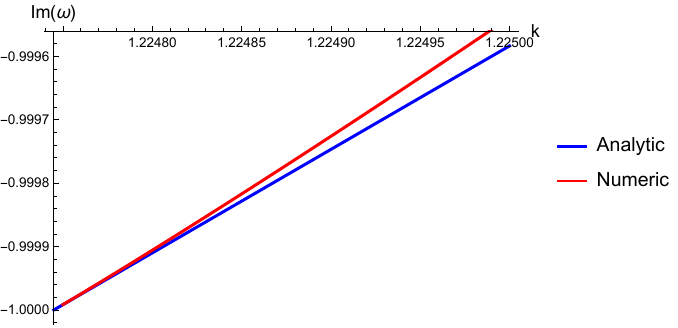}
  \caption{$m=\frac{3}{2}$}
  \label{subfig:: m=1.5}
\end{subfigure}
\caption{Comparison between analytic and numerical results for the emerging pole
in two cases with positive $m$.}
\label{fig:: Positive m plots}
\end{figure}

\paragraph{$m>1$: linear behaviour}
For these values of $m$, the leading term in the expansion~\eqref{eq:: 2F1 expansion} is the linear one. Thus:
\begin{equation}
    P(\veps,\delta k) \simeq \frac{i\veps}{\tau_*} \Big(1+\frac{1}{1-m}\Big) +2i\tau_*^{m/2} \sqrt{\frac{c}{m}} \delta k \;, 
\end{equation}
that gives us the linear expression:
\begin{equation}\label{eq:: Linear epsilon behaviour}
    \veps(k) \simeq \frac{2(m-1)}{2-m} \sqrt{\frac{c\tau_*^{2+m}}{m}}  (k-k_c)\;.
\end{equation}
Within the exact power-law family the coefficient in~\eqref{eq:: Linear epsilon behaviour} changes sign at $m=2$. For $m>2$ the solution lies on the $k<k_c$ side. Section~\ref{Sec:: Breakdown for general m} identifies it with the diffusive branch and gives
\begin{equation}\label{eq:: termination momentum}
    k_{\mathrm{term}}=k_c=\sqrt{\frac{m}{c\tau_*^{2+m}}}\;,
    \qquad m\geq2\;.
\end{equation}
\paragraph{$0<m<1$: non-linear behaviour}
When $m$ is lower than $1$, the term $\veps^m$ becomes more important than the linear one. Therefore, by solving again $P(\veps,\delta k)=0$, we obtain, at leading order:
\begin{equation}
    \veps(k) \simeq\Bigg(\frac{2\sqrt{c\tau_*^{2+m}}\sin(\pi m)}{\pi m^{3/2}}(k-k_c)\Bigg)^{1/m}\;.
\end{equation}
In Figure~\ref{fig:: Positive m plots} we can appreciate the difference between two explicit cases of linear and non-linear behaviour.

\paragraph{$-1<m<0$: divergence at the branch point}
For negative values of $m$, $B(\omega)$ diverges at the branch point. As we can see from~\eqref{eq:: 2F1 expansion}, the leading term is $\veps^m$ and the master equation~\eqref{Eq:Branch cut modes equation} becomes:
\begin{equation}
    1+\frac{\pi ck^2\tau_*^{2+m}}{\sin(\pi m)} \veps^m \simeq0 \qquad \Longrightarrow  \qquad \veps(k) \simeq \Bigg(\frac{\pi c\tau_*^{2+m}}{|\sin(\pi m)|} k^2\Bigg)^{1/|m|}\;.
\end{equation}
It is interesting to inspect the residue of the charge-charge correlator at this pole. Indeed, by repeating our discussion of Section~\ref{Sec:: Log branch cut}, we find that:
\begin{equation}
    \text{Res}_{\omega_m}[\langle J^0 J^0\rangle] \propto k^{2/|m|}\;,
\end{equation}
which is always subleading with respect to the hydrodynamic pole, though only in a perturbative way, unlike what we had with the logarithmic branch point.

\paragraph{Integer $m$}
For positive integer $m=n$, the corresponding limiting forms of the same continuation formulas give
\begin{subequations}
    \begin{align}
    \prescript{}{2}{F}_1(1,1;3;1-\veps) &= 2+2\veps(\log\veps+1)+O(\veps^2\log\veps) \\
    \prescript{}{2}{F}_1(1,1;2+n;1-\veps)
&= \frac{n+1}{n}
\Big(1-\frac{\veps}{n-1}\Big)
+O(\veps^2)+O(\veps^n\log\veps)\;,
\qquad n\geq2\;.
\end{align}
\end{subequations}

We start by studying the $n=1$ case; the mode equation~\eqref{Eq:Branch cut modes equation} around the branch point becomes:
\begin{equation}
    P(\veps,\delta k) \simeq i c k_c \tau_*^2 \Big( k_c \veps \log\veps + 2\delta k\Big) =0\;,
\end{equation}
that is solved by the Lambert $W$ function\footnote{The branch of the $W$ function can be selected with the same discussion of Section~\ref{Sec:: Log branch cut}.}, namely:
\begin{equation}
    \veps(k) = \texp{W_{-1}\Big(-\frac{2\delta k}{k_c}\Big)} \simeq -\frac{2\delta k}{k_c \log(\frac{2\delta k}{k_c})}\;.
\end{equation}
We recall that $m=1$ is the threshold for linear behaviour: indeed, we find a logarithmic correction.

For $n\geq3$ we can repeat the same procedure and obtain the same expression as~\eqref{eq:: Linear epsilon behaviour}, with the sign discussed above. The value $n=2$ is genuinely marginal and must be treated separately, as is already apparent from the pole of~\eqref{eq:: Linear epsilon behaviour} at $m=2$. Indeed, using $\prescript{}{2}{F}_1(1,1;4;1-\veps)
=\tfrac{3}{2}(1-\veps)+O(\veps^2\log\veps)$ and $ck_c^2\tau_*^3=2$, the two leading terms of the mode equation organize into
\begin{equation}
    P(\veps,\delta k) \simeq \frac{i}{\tau_*}\,(1-\veps)\,\frac{2\delta k}{k_c}\;,
\end{equation}
so that the dependence on $\veps$ cancels identically at this order: at $m=2$ the emergence of the pole is not controlled by the leading endpoint behaviour, and one has to go to the subleading $\veps^2\log\veps$ term. This cancellation is a special feature of the pure-power family. The global analysis of Section~\ref{Sec:: Breakdown for general m} shows that, within this family, it coincides with the point at which the interior double root reaches the endpoint.

These formulas realize the local classes of Table~\ref{tab:: Summary Branch points}. Their coefficients and root identifications are specific to the pure-power density~\eqref{eq:: Exact power law density}.

\paragraph{Continuation across the branch point}
For $0<m<1$ the continuation also shows explicitly how the physical-sheet pole connects through the branch point to the hydrodynamic pole on a non-principal sheet. For $m=1/2$:
\begin{equation}
    e^{in\pi} \sqrt{\veps(k)} = \frac{4}{\pi}\sqrt{2 c\tau_*^{5/2}} (k-k_c)\;,
\end{equation}
where $n$ labels the sheets. We see that the equation is consistent only if $k<k_c$ for $n$ odd, i.e. the pole gets closer to the branch point, and if $k>k_c$ for $n$ even, i.e. the pole has changed sheet and goes away from the branch point.

\subsubsection{Real momentum termination within the power-law family}\label{Sec:: Breakdown for general m}
For the fully specified density~\eqref{eq:: Exact power law density}, the global spectral curve can be solved in closed parametric form. As long as the poles sit on the imaginary axis we may set $\omega=-iy/\tau_*$, so that $i\omega\tau_*=y$ with $y\in(0,1)$, the endpoint corresponding to $y=1$. The master equation~\eqref{Eq:Branch cut modes equation} with the algebraic density then inverts exactly for the momentum,
\begin{equation}\label{eq:: spectral curve general m}
    c\,\tau_*^{2+m}\,k^2 = h_m(y)\equiv \frac{(1+m)\,y}{\prescript{}{2}{F}_1(1,1;2+m;y)}\;,
\end{equation}
which is the analogue of the spectral curve $q(z)$ of Section~\ref{Sec:: Log branch cut}, to which it reduces for $m\to0$ upon using $\prescript{}{2}{F}_1(1,1;2;y)=-\log(1-y)/y$. Poles of the retarded correlator on the imaginary axis are the preimages of $k^2$ under $h_m$, and two of them merge exactly where $h_m'(y)=0$, i.e. at an interior maximum of $h_m$. This is the same double-root condition used in Section~\ref{Sec:: Log branch cut}, now solved for arbitrary endpoint exponent.

The structure of $h_m$ on $(0,1)$ is the following. One has $h_m(1)=m$ for $m>0$, which reproduces the threshold $k_c=\sqrt{m/(c\tau_*^{2+m})}$ obtained above from Gauss' formula. For $m<2$ the function $h_m$ possesses a unique interior maximum at some
$y_*(m)<1$, while for $m\geq2$ it is monotonically increasing on $(0,1)$,
with its maximum attained at $y=1$.\footnote{Writing
$F_m(y)={}_2F_1(1,1;2+m;y)$, the sign of $h_m'$ is the sign of
$g_m=F_m-yF_m'$, with $g_m'=-yF_m''<0$. Moreover,
$g_m(1)=(m+1)(m-2)/[m(m-1)]$ for $m>1$, while
$g_m(1^-)<0$ for $m\leq1$.} Consequently:
\begin{itemize}
    \item[-] For $m<2$ the hydrodynamic branch is terminated by a \emph{pole-pole collision} at
    \begin{equation}\label{eq:: k break}
        k_{\mathrm{term}}=\frac{\sqrt{h_m(y_*)}}{\sqrt{c}\,\tau_*^{1+m/2}}\;, \qquad \omega_{\mathrm{term}}=-\frac{i\,y_*(m)}{\tau_*}\;,
    \end{equation}
    which, by the same expansion of $P$ around the double root performed in Section~\ref{Sec:: Log branch cut}, is again a square-root branch point of $\omega(k)$ in the complex momentum plane. For $0<m<2$ one has $k_{\mathrm{term}}>k_c$: the pole first emerges from the branch point at $k_c$ and only later collides with the diffusive one. For $-1<m\leq0$ the emergent pole exists at arbitrarily small momentum and the collision is the only obstruction encountered along the positive real momentum axis.
    \item[-] For $m\geq2$ no interior maximum exists, no pole is emitted, and the diffusive pole runs into the branch point exactly at $k_c$, so that $k_{\mathrm{term}}=k_c$ as anticipated in~\eqref{eq:: termination momentum}.
\end{itemize}
The transition between the two regimes is continuous: $y_*(m)\to1$ and $k_{\mathrm{term}}\to k_c$ as $m\to2^-$. The marginality of $m=2$ found in the expansion around the branch point is therefore not an artefact: it is the value at which the collision point reaches the endpoint of the continuum and the pole-pole mechanism degenerates into the pole/branch-point one.

\begin{table}[t]
    \centering
    \begin{tabular}{c|c|c|c|c}
        $m$ & $y_*$ & $\sqrt{c}\,\tau_*^{1+m/2}k_{\mathrm{term}}$ & $\sqrt{c}\,\tau_*^{1+m/2}k_{c}$ & $k_{\mathrm{term}}/k_c$\\
        \hline
        $-0.75$ & $0.5598$ & $0.2702$ & - & -\\
        $-0.5$ & $0.6151$ & $0.4074$ & - & -\\
        $-0.25$ & $0.6667$ & $0.5270$ & - & -\\
        $0$ & $0.7153$ & $0.6382$ & $0$ & $\infty$\\
        $0.5$ & $0.8045$ & $0.8467$ & $0.7071$ & $1.197$\\
        $1$ & $0.8834$ & $1.0437$ & $1.0000$ & $1.044$\\
        $1.5$ & $0.9512$ & $1.2327$ & $1.2247$ & $1.007$\\
        $1.9$ & $0.9933$ & $1.3786$ & $1.3784$ & $1.0002$\\
        $\geq2$ & $1$ & $\sqrt{m}$ & $\sqrt{m}$ & $1$\\
    \end{tabular}
    \caption{Real momentum termination of the hydrodynamic branch within the exact power-law family~\eqref{eq:: Exact power law density}, obtained from the interior maximum of the spectral curve~\eqref{eq:: spectral curve general m}. The row $m=0$ reproduces the closed-form logarithmic result $\sqrt{c}\tau_*k_c=0.6381727$ of Section~\ref{Sec:: Log branch cut}, which provides a non-trivial check of the two independent computations.}
    \label{tab:: Breakdown momenta}
\end{table}

Table~\ref{tab:: Breakdown momenta}  summarizes the exact pure-power result. The transition at $m=2$ is specific to the pure-power family. For a general density, changing the holomorphic part of $B$ can move, create, or remove interior extrema of the spectral curve.

\subsection{Higher-order gradients in the non-hydrodynamic sector}\label{Sec:: Gradient corrections}
We now  consider three representative higher-gradient deformations within the EFT regime. First consider corrections local in the relaxation label,
\begin{equation}
    X(\tau,\tau';k)=\chi_v\big[1+k^2\ell^2(\tau)+O(k^4)\big]\delta(\tau-\tau')\;,
\end{equation}
together with smooth momentum corrections to $\alpha$ and $\gamma_+$. Each constituent then relaxes at
\begin{equation}
    \Gamma(\tau,k)=\frac{1+k^2\ell^2(\tau)+O(k^4)}{\tau}\;.
\end{equation}
Using the dressed relaxation time $T=1/\Gamma$ as integration variable preserves the Stieltjes form. If $\tau\mapsto T$ is a local diffeomorphism at the endpoint, its Jacobian is smooth and non-zero there, so
\begin{equation}
    F(T,k)\sim\big[T_*(k)-T\big]^m
\end{equation}
with the same exponent $m$. The branch point moves and the analytic coefficients are renormalized, but the local scalings in Table~\ref{tab:: Summary Branch points} are unchanged. In the logarithmic class, for example, the isolated pole remains separated from the moving endpoint by an essential singularity,
\begin{equation}
    \omega_b(k)-\omega_*(k)\sim
    \exp\!\left[-\frac{A(k)}{k^2}\right]\;,
\end{equation}
provided the corrected retarded germ continues to support that root. Any collision generally moves and remains a global property of the complete kernel.

Second, a non-local correction $k^2Y(\tau,\tau')$ can mix relaxation labels and may generate discrete edge states. In a weak rank-one edge problem, the relevant integral has the schematic form
\begin{equation}
    \int_{\Gamma_{\min}}\dif\Gamma\;
    \frac{\rho(\Gamma)}{\Gamma-\Gamma_{\min}+\delta}\;,
    \qquad \rho(\Gamma)\sim(\Gamma-\Gamma_{\min})^m\;.
\end{equation}
Its inversion reproduces the exponential scale for $m=0$ and the power $k^{2/|m|}$ for $-1<m<0$, preserving the local edge exponents without fixing the global trajectory.

Finally, gradient terms may endow the continuum with a finite group velocity and spread the cut away from the imaginary axis, as in kinetic streaming. Stability then constrains the full finite-momentum spectrum, and the endpoint may become momentum dependent. The near-edge pole and real-momentum termination consequently depend on the complete finite-momentum kernel~\cite{Romatschke:2015gic,Kurkela:2017xis,Bajec:2024rta,Brants:2024kin}. 

\section{Multiple branch cuts}\label{Sec:: Multiple branch cuts}
We now generalize Section~\ref{Sec:: Single branch cut} to several coupled continua, allowing cuts away from the imaginary axis. We promote
\begin{equation}
v^\mu(\tau)\longrightarrow \vec v^{\,\mu}(\vec\tau),
\end{equation}
so that $X$ and $\Sigma$ become matrix-valued integral kernels and $\vec\gamma_\pm$ vector-valued functions of the continuous labels. The SK action retains the form of Section~\ref{Sec:: Single branch cut}. Upon taking the components of $\vec\tau$ to have common support, $\text{KMS}_\Theta$ symmetry subsequently requires
\begin{equation}\label{eq:: Branch cuts KMS conditions}
    X(\vec{\tau},\vec{\tau}')=\tilde{X}^t(\vec{\tau}',\vec{\tau}) \;, \quad \Sigma(\vec{\tau},\vec{\tau}')=\tilde{\Sigma}^t(\vec{\tau}',\vec{\tau})\;, \quad \gamma_\pm(\vec{\tau}) = \pm\tilde{\gamma}_\pm(\vec{\tau})\;.
\end{equation}
As frame conditions we exactly recover~\eqref{eq:: Branch cut frame conditions}. For brevity, first take $\Theta$ not to include charge conjugation, so $\gamma_-=0$ and the relevant noise kernel is symmetric. Positivity of $\mathrm{Im} \; S$ requires
\begin{equation}
    \sigma\geq0\;,
    \qquad \Sigma\geq0\;.
\end{equation}
For a positive semidefinite $\Sigma$, the generalized Schur-complement condition is
\begin{equation}\label{eq:: Schur bound}
    |\gamma_+\rangle\in\overline{\operatorname{Ran}\Sigma}\;,
    \qquad
    \sigma-\langle\gamma_+|\Sigma^+|\gamma_+\rangle\geq0\;,
\end{equation}
where $\operatorname{Ran}\Sigma$ denotes the range (image) of the operator $\Sigma$, and $\Sigma^+$ is the generalized inverse on this dynamical range. The second frame condition in~\eqref{eq:: Branch cut frame conditions} saturates this bound. The full noise matrix is therefore positive semidefinite and has a null direction: in the MC frame the stochastic source carried by the current is not independent of that in the relaxing sector. Allowing an additional dissipative contribution directly in $\Jcal^\mu$ would move the theory into the interior of the positive cone without changing the analytic discussion below.

It is immediate to verify that the equations of motion generated by this action are solved by~\eqref{Eq:Responses}, with the same definition of $B(\omega)$. At zero momentum, the continuum is supported at those frequencies
for which the operator
\begin{equation}
    A(\omega)\equiv X-i\omega\Sigma
\end{equation}
ceases to be invertible, or equivalently
\begin{equation}
    0\in\operatorname{spec}\!\left[X-i\omega\Sigma\right]\;.
\end{equation}

\subsection{Reality and stability of the branch cuts}\label{Sec:: Reality of the branch cuts}

Equivalently, the zero-momentum continuum is determined by the generalized
eigenvalue problem
\begin{equation}
    X|f\rangle=i\omega\,\Sigma|f\rangle\;.
\end{equation}
When charge conjugation is absent, KMS symmetry makes $X$ and $\Sigma$
symmetric. On the subspace where $\Sigma$ is positive, this equation is
equivalent to an ordinary eigenvalue problem for the symmetric operator
$\Sigma^{-1/2}X\Sigma^{-1/2}$, and therefore
$i\omega\in\mathbbm{R}$. This conclusion fixes the axis but not the sign. With the convention $e^{-i\omega t}$, stability is the additional requirement $\operatorname{Im}\omega\leq0$, or $i\omega\geq0$, on the generalized relaxation spectrum. Thus KMS symmetry together with noise positivity places the zero-momentum cuts on the imaginary axis, while positivity of the relaxation rates selects the lower half-plane.

To obtain zero-momentum branch points with non-zero real part in this construction, the dynamical KMS transformation must be chosen as $\Theta=CT$ rather than $\Theta=T$, which permits the antisymmetric structures used below. Reality of retarded correlators then requires off-axis branch points to occur in pairs $(\omega,-\omega^*)$.

\subsection{Two branches}
Let us consider an explicit example of $N=2$ vectors described by the same continuous parameter:
\begin{equation}
    \vec{v}^\mu(\tau) = \begin{pmatrix}
        v_1^\mu(\tau) \\
        v_2^\mu(\tau)
    \end{pmatrix}\;.
\end{equation}
We consider the operators:
\begin{equation}
    X(\tau,\tau') =\chi_v \delta(\tau-\tau')\begin{bmatrix}
        x_1 & x_m  \\
        x_m & x_2
    \end{bmatrix} \;, \qquad \Sigma(\tau,\tau') =\chi_v \delta(\tau-\tau')\begin{bmatrix}
        \tau & \eta(\tau) + \xi(\tau) \\
        \eta(\tau) - \xi(\tau) & \lambda \tau
    \end{bmatrix}\;,
\end{equation}
where $x_i$ and $\lambda$ are numbers. For $\Theta=CT$, the KMS condition (4.1) relates the kernel at opposite chemical potentials,
\begin{equation}
\Sigma(\mu)=\Sigma^{t}(-\mu).
\end{equation}
It therefore requires
\begin{equation}
\eta(-\mu)=\eta(\mu),
\qquad
\xi(-\mu)=-\xi(\mu).
\end{equation}
Thus the symmetric off-diagonal structure is even in $\mu$, while the
antisymmetric one is allowed provided it is odd in $\mu$. With these expressions, we can easily find $\omega(\tau)$ that solves $\det[A]=0$:
\begin{equation}\label{eq:: Two branch cuts explicit}
    \omega_\pm (\tau) =\frac{i}{2(\eta^2(\tau)-\xi^2(\tau)-\lambda \tau^2)} \Bigg[ \tau(\lambda x_1 + x_2) - 2\eta(\tau) x_m \pm \sqrt{\Delta(\tau)} \Bigg]\;,
\end{equation}
where
\begin{equation}
    \Delta(\tau) =\Big(\tau(\lambda x_1 + x_2)-2\eta(\tau) x_m\Big)^2 - 4(x_m^2-x_1x_2)\Big(\eta^2(\tau)-\xi^2(\tau) - \lambda \tau^2\Big)\;.
\end{equation}
Let us consider $\chi_v>0$ and $\tau\in[\tau_{min},\tau_{\max}]$ positive; the positivity of $\Sigma$,
\begin{equation}
    \int_{\tau_{min}}^{\tau_{max}} \dif\tau \chi_v \vec{v}^t(\tau) \begin{bmatrix}
        \tau & \eta(\tau)+\xi(\tau) \\
        \eta(\tau)-\xi(\tau) & \lambda \tau
    \end{bmatrix} \vec{v}(\tau)\geq0\;,
\end{equation}
tells us that:
\begin{equation} \label{Eq: Condition for positiveness of Sigma}
    \lambda\geq0 \;, \qquad \lambda \tau^2 \geq \eta^2(\tau)\;.
\end{equation}
Note that the antisymmetric part of $\Sigma$ drops out of these conditions. In particular, for a real antisymmetric kernel $A$, $v^\dagger A v$ is purely imaginary. Consequently, the term $i\,v^\dagger A v$ is real and does not contribute to $\operatorname{Im}S$. Positivity therefore constrains only the symmetric part of $\Sigma$. The function $\xi(\tau)$ is therefore unconstrained by unitarity, which is precisely what makes the off-axis branch cuts constructed below possible.

\begin{figure}
    \centering
    \includegraphics[width=0.5\linewidth]{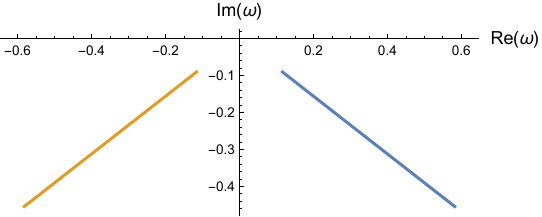}
    \caption{Two branch cuts~\eqref{eq:: Two branch cuts explicit} with $x_1=x_2=2,\; x_m=1,\; \lambda=1.5, \;\eta=0, \;\xi(\tau)=2\tau$ and $\tau\in[0,5]$.}
    \label{fig:: Two branches}
\end{figure}

For brevity, we now consider a specific case $\eta(\tau)=0$ and $\xi(\tau) = a\tau$, with $a(-\mu)=-a(\mu)$. Then, the two branch cuts~\eqref{eq:: Two branch cuts explicit} are stable and have a non-zero real part if:
\begin{subequations}
    \begin{align}
        &\lambda x_1 + x_2 >0\\
        &\det X \propto x_1 x_2 - x_m^2 > 0 \\
        &a^2 > \frac{(\lambda x_1 - x_2)^2 +4\lambda x_m^2}{4(x_1x_2-x_m^2)}\;.
    \end{align}
\end{subequations}
An explicit plot of the two branches satisfying these constraints is shown in Figure~\ref{fig:: Two branches}.

Reality, $G_R(\omega)^*=G_R(-\omega^*)$, pairs the off-axis branch points and their local discontinuity data, as realized by $i\omega_+=(i\omega_-)^*$ in~\eqref{eq:: Two branch cuts explicit}. The cut paths themselves remain conventional: a reflection-symmetric choice makes the reality relation manifest, while an asymmetric choice represents the same continuation. In the present parametrization both cuts arise from the same $\tau$ contour and therefore move together under its deformation. This correlation is parametrization-dependent, not intrinsic analytic data .

\subsection{Multiple emergent poles}
We now ask how the near-edge poles of Section~\ref{Sec:: Single branch cut}  generalize when several cuts are present. First of all, let us consider the explicit case:
\begin{equation}
    X(\tau,\tau') =\chi_v \delta(\tau-\tau') \bar{X} \;, \qquad \Sigma(\tau,\tau') =\chi_v \tau \delta(\tau-\tau') \bar{\Sigma} \;,
\end{equation}
where $\bar X$ and $\bar\Sigma$ are matrices independent of $\tau$ and are taken to be non-singular on the coupled subspace. In addition we impose:
\begin{equation}
    \vec{\alpha}(\tau) = \vec{\alpha} f_\alpha(\tau) \;, \qquad \vec{\gamma}_+(\tau) = \vec{\gamma}_+ f_\gamma(\tau)\;.
\end{equation}
Frame constraint~\eqref{eq:: Branch cut gamma alpha frame constraint} becomes:
\begin{equation}
    \frac{f_\gamma(\tau)}{\tau} \vec{\gamma}_+^t \bar{\Sigma}^{-1} \bar{X} = \vec{\alpha}^t f_\alpha(\tau)\;.
\end{equation}
We can substitute this expression into the definition of $B(\omega)$:
\begin{equation}
    B(\omega) = \int \dif\tau \; \vec{\gamma}_+^t (\bar{X}-i\omega \tau \bar{\Sigma})^{-1} (\bar{\Sigma}^{-1}\bar{X})^t \vec{\gamma}_+ \frac{f_\gamma^2(\tau)}{\tau}\;.
\end{equation}

As an instructive example, we explicitly consider $N=2$ and choose $f_\gamma(\tau)\propto\sqrt{\tau}$, in analogy with~\eqref{eq:: Log branch cut alpha gamma choices}, in order to obtain logarithmic branch points. Indeed, we obtain:
\begin{equation}
    B(\omega) = \frac{1}{\omega} \Bigg[r_+\log \Big(1-\frac{\omega}{\omega_+}\Big) + r_-\log \Big(1-\frac{\omega}{\omega_-}\Big)\Bigg]\;.
\end{equation}
Here, $\omega_\pm$ are the two branch points, satisfying $i\omega_+=(i\omega_-)^*$. Near each logarithmic branch point,  Section~\ref{Sec:: Local endpoint emergence} allows an exponentially close root when its phase places it on the retarded sheet. The pair is related by $\omega\to-\omega^*$. Their later trajectories and collisions depend on the complete kernel, so no  universal global pattern follows from the branch points alone.

\section{A microscopic realization: energy-dependent relaxation times}\label{Sec:: Microscopic realization}
A simple microscopic generator of the Stieltjes kernel is the energy-diagonal relaxation-time approximation (RTA). In this approximation a quasiparticle perturbation at energy $\epsilon$ relaxes independently with time $\tau(\epsilon)$. At zero momentum the optical conductivity is~\cite{Ziman1960}
\begin{equation}\label{eq:: Boltzmann conductivity}
    \sigma(\omega)=\int\dif\epsilon\;w(\epsilon)
    \frac{\tau(\epsilon)}{1-i\omega\tau(\epsilon)}\;,
    \qquad
    w(\epsilon)=\left(-\frac{\partial n_0}{\partial\epsilon}\right)
    g(\epsilon)\frac{v(\epsilon)^2}{d}\;.
\end{equation}
For real $\omega$, \eqref{eq:: Boltzmann conductivity} is the physical optical conductivity, understood as the boundary value of the retarded response. Upon analytic continuation into complex frequency, each relaxation time contributes a singularity at $\omega=-i/\tau(\epsilon)$. When the relaxation times form a continuum, these singularities condense into the branch cut represented by the Stieltjes transform which has the form~\eqref{eq:: B Stieltjes representation}. In the case that $\tau(\epsilon)$ is not one-to-one, the density must be taken as the sum
\begin{equation}\label{eq:: RTA spectral density}
    f(\tau)=\chi_v\,\tau
    \sum_{\epsilon_i:\,\tau(\epsilon_i)=\tau}
    w(\epsilon_i)\left|\frac{\dif\epsilon_i}{\dif\tau}\right|\;,
\end{equation}
up to the normalization convention for $\chi_v$. The factor of $\tau$ follows from the numerator of~\eqref{eq:: Boltzmann conductivity}. It is finite and non-zero at a gapped upper endpoint, so it does not change the endpoint exponent there.

Suppose $\tau(\epsilon)$ has a smooth interior maximum,
\begin{equation}
    \tau(\epsilon)=\tau_*-a(\epsilon-\epsilon_0)^2+\cdots\;,
    \qquad w(\epsilon_0)\neq0\;.
\end{equation}
The two inverse branches give a Jacobian $(\tau_*-\tau)^{-1/2}$, hence $m=-1/2$ and $\delta\omega\sim k^4$ near the endpoint. On the other hand, a logarithmic endpoint, $m=0$, is obtained when the maximum of $\tau$ occurs at an edge of a bounded energy interval with non-zero slope and with $w$ finite and non-zero. If the weight vanishes at the same edge, its vanishing order shifts $m$ to a positive value. Conversely, an unbounded relaxation-time distribution can make the cut reach $\omega=0$ rather than produce a gapped branch point. Equation~\eqref{eq:: RTA spectral density} therefore provides a direct microscopic interpretation of the endpoint expansion~\eqref{eq:: General endpoint expansion} within the energy-diagonal RTA.

This realization is specific to the energy-diagonal RTA. A general linearized collision term is an integral operator whose eigenfunctions need not be energy eigenstates and whose spectral measure need not be~\eqref{eq:: RTA spectral density}; the limitations of energy-dependent RTA are discussed in~\cite{Hu:2024rta}. At finite momentum, streaming further modifies the kernel and can spread the cut in the real-frequency direction. Matching the finite-momentum effective action therefore requires the spectrum and overlaps of the full streaming-plus-collision operator, rather than the zero-momentum density alone.

Relaxation-time distributions are familiar in dielectric and glassy response \cite{ColeCole1941,HavriliakNegami1967}. Here the SK construction additionally supplies the statistical sector and identifies the finite-momentum data required for pole trajectories.

\section{Conclusions and outlook}\label{Sec:: Conclusions}

We have constructed Gaussian Schwinger-Keldysh effective actions for a conserved charge coupled to discrete or continuous families of relaxing vector fields. For finitely many modes, the conductivity is a sum of Drude-like terms weighted by their overlaps with the physical current.  Taking the relaxation spectrum to a continuum produces branch cuts and, for the relaxation-diagonal class studied here, a Stieltjes kernel determined by a relaxation-time density. The construction supplies both the retarded response and its KMS-related fluctuation sector.

This formulation clarifies which properties of a non-hydrodynamic continuum are local and which require its complete spectral measure. Dynamical KMS symmetry, positivity and stability constrain the zero-momentum relaxation spectrum, while retarded analyticity fixes the physical sheet and separates intrinsic branch-point data from the conventional placement of cuts. The endpoint behaviour of the spectral density determines the local branch singularity and near-edge scaling, but not the complete pole trajectory. In the uniform example, a pole and logarithmic endpoint have identical perturbative dispersions but are separated beyond all orders before the pole collides with diffusion at finite real momentum. In the exact power-law family, this pole-pole collision moves continuously to the continuum endpoint as $m\to2$. These real-momentum singularities constrain, but need not by themselves determine, the radius of convergence in the full complex momentum plane.

Smooth local gradient corrections move the continuum endpoint while preserving its exponent and associated near-edge scaling. More general mixing and propagation supply genuinely new finite-momentum data. The energy-dependent relaxation-time approximation provides a direct zero-momentum microscopic realization of the Stieltjes kernel, but generic kinetic matching requires the spectrum and current overlaps of the full streaming-plus-collision operator.

Several directions seem worth pursuing. Microscopic matching beyond energy-diagonal RTA should extract the spectral measure and current overlaps of the full streaming-plus-collision operator, connecting the present construction to kinetic onset transitions~
\cite{Romatschke:2015gic,Kurkela:2017xis,Bajec:2024rta,Brants:2024kin}, kinetic-theory relaxation terms~
\cite{Amoretti:2023hbr}, shift-symmetry EFTs~\cite{An:2025shift}, and the reorganization of holographic quasinormal modes into cuts~\cite{Grozdanov:2016vgg}. Beyond Gaussian order, hydrodynamic interactions generate long-time tails and fluctuation-induced cuts~\cite{ChenLin2018,Delacretaz:2020nit}. Their interplay with a pre-existing non-hydrodynamic continuum and the non-local statistical action obtained after integrating that continuum out, are natural extensions. Finally, coupling the relaxing continuum to momentum and energy would extend the construction to stress-tensor correlators and the sound and shear channels \cite{Amoretti:2021fch,Amoretti:2021lll,Amoretti:2023vhe}, with possible applications to the quark-gluon plasma and anomalous optical response.

Taken together, these results provide a Schwinger-Keldysh framework in which discrete relaxation modes, non-hydrodynamic continua and their associated analytic structures can be treated within a common effective description.

\appendix

\section{Explicit heavy modes integration in the SK action}\label{App: Integration of heavy modes}
We illustrate the heavy-mode reduction with an exact Maxwell-Cattaneo example and a perturbative $N=2$ example.

\subsection{Exact computation: standard MC model}
For the standard MC model the reduction can be performed exactly; let us explicitly show the process of integrating out a rapidly decaying non-hydrodynamic mode. The SK action is:
\begin{align}\label{eq:: Action for standard MC}
    S_{MC}[B_{s\mu},v_s^\mu] =& \xint \Bigg\{ n B_{a\mu}u^\mu -\frac{\tau}{\sigma} \Jcal_{a\mu} \Jcal^\mu_r + \nonumber \\
    &+iT\Delta^{\mu\nu}\left(\sigma B_{a\mu} +\tau \Jcal_{a\mu} \right) \left[B_{a\nu} + i \beta^\rho \partial_\rho B_{r\nu} + \frac{\tau}{\sigma}\left( \Jcal_{a\nu}+i\beta^\rho\partial_\rho \Jcal_{r\nu} \right)\right] \Bigg\} \;.
\end{align} 
In the fluid rest frame, the deterministic equations of motion are:
\begin{align}
    &\partial_t n + \partial_i \Jcal^i =0 \\
    &\Jcal^i + \tau \partial_t \Jcal^i = \sigma E^i -D\partial^in\;.
\label{eq:: MC deterministic eoms}
\end{align}
In Fourier space, the longitudinal sector becomes:
\begin{equation}
    \omega \begin{pmatrix}
        \frac{n}{\sqrt{\tau}} \\
        \frac{\Jcal^{||}}{\sqrt{D}}
    \end{pmatrix}= \mathsf{W}\begin{pmatrix}
        \frac{n}{\sqrt{\tau}} \\
        \frac{\Jcal^{||}}{\sqrt{D}}
    \end{pmatrix} + \mathsf{T}^{-1}\begin{pmatrix}
        0 \\
        \frac{i\sigma E^{||}}{\sqrt{\tau D}}
    \end{pmatrix}
\end{equation}
where
\begin{align}
    \mathsf{W} &= \begin{bmatrix}
        0 & \frac{\sqrt{D}\,k}{\sqrt{\tau}} \\
        \frac{\sqrt{D}\,k}{\sqrt{\tau}} & -\frac{i}{\tau}
    \end{bmatrix}\;, \\
    \mathsf{T}&=\begin{bmatrix}
        \sqrt{D}^{-1} & 0 \\
        0& \sqrt{\tau}
    \end{bmatrix}\;.
\end{align}
One can easily verify that the eigenvalues of $\mathsf{W}$ are the standard MC poles
\begin{equation}
    \omega_\pm = \frac{i}{2\tau}\left(\pm\sqrt{1-4Dk^2\tau}-1\right)\;.
\label{eq:: MC poles appendix}
\end{equation}
 The eigenvectors are functions of the momentum; up to $o(k^2)$, one obtains:
\begin{equation}
    \psi_l=\begin{pmatrix}
        1+\frac{D\tau k^2}{2} \\
        -i\sqrt{D\tau}k
    \end{pmatrix}
    \;, \qquad\psi_h=\begin{pmatrix}
        i\sqrt{D\tau}k \\
        1+\frac{D\tau k^2}{2}
    \end{pmatrix}\;.
\end{equation}
At $k=0$, $n$ is diffusive while $\Jcal^{||}$ and  $\Jcal^i_\perp$ are gapped. We therefore use the electric-field variables
\begin{align}
    B_{||}(p) &=-\frac{i}{\omega} E_{||}(p) - \frac{k}{\omega} B_0(\psi_l(p),\psi_h(p))  \\
     B_\perp^i (p) &= -\frac{i}{\omega} E_\perp^i(p)\;.
\label{eq:: E field change of variables}
\end{align}
Because the theory is Gaussian, integrating out $\psi_h$ is equivalent to its saddle-point equation,
\begin{equation}
    \psi_h^a(\omega,k) =\psi^a_h(\psi_l^a,E_{||}^a;\omega,k)\; ,
\end{equation}
whose small-$k$ form is
 \begin{eqnarray}
    \psi_h^a(\omega,k) = -\frac{\sigma E^a_{||}(\omega,k)}{\sqrt{D}(1+i\omega\tau)} +ik\sqrt{D\tau} \psi^a_l(\omega,k)\;.
\end{eqnarray}
Substitution gives
\begin{align}
    S_\text{eff}[\psi_l^s, E_i^s] = \pint\; \Bigg\{&
    \begin{pmatrix}
        \psi_l^a(-p) & E_{||}^a(-p) 
    \end{pmatrix} \begin{bmatrix}
        G_{ll}^R(p) & G_{l||}^R(p) \\
        G_{||l}^R(p) & G_{||,||}^R(p)
    \end{bmatrix}\begin{pmatrix}
        \psi_l^r(p) \\
        E_{||}^r(p)
    \end{pmatrix} + \nonumber \\
    &+\begin{pmatrix}
        \psi_l^a(-p) & E_{||}^a(-p) 
    \end{pmatrix} \begin{bmatrix}
        G_{ll}^S(p) & G_{l||}^S(p) \\
        G_{||l}^S(p) & G_{||,||}^S(p)
    \end{bmatrix}\begin{pmatrix}
        \psi_l^a(p) \\
        E_{||}^a(p)
    \end{pmatrix}
    + \nonumber \\
    &+ E_{\perp,i}^a(-p) G^R_{\perp\perp}(p) E^i_{r\perp}(p)+ E_{\perp,i}^a(-p) G^S_{\perp\perp}(p) E^i_{a\perp}(p)\Bigg\} \;.
\label{eq:: MC effective action}
\end{align}

Gauge invariance gives
\begin{align}
    \left( \frac{\delta}{\delta A_{||}^r(p)}\frac{\delta}{\delta A_{0}^a(-p)} - \frac{\delta}{\delta A_{0}^r(p)}\frac{\delta}{\delta A_{||}^a(-p)}\right) S_\text{eff} &= \nonumber\\
    =C_{ln} (-p) G^R_{l||}(p) + C_{ln}(p) G^R_{||l}(p) &=0\;,
\end{align}
where $\psi_l(p)=(C_{ln}(p),C_{lj}(p))^t$. Gaussian integration also preserves dynamical KMS with the normalization of~\eqref{eq:: MC effective action},
 \begin{equation}\label{eq:: FDT effective kernels}
    \mathbb{G}^{S}(p)
    =
    \frac{T}{2\omega}
    \left[
        \mathbb{G}^{R}(p)-\mathbb{G}^{A}(p)
    \right]\;,
\end{equation}
where the advanced kernel includes the appropriate time-reversal transformation. The Gaussian Schur complement therefore inherits the same KMS relation.

We now consider the conductivity as defined in~\cite{Amoretti:2025kem}:
\begin{equation}\label{eq:: conductivity definition}
    \sigma^{ij}(\omega,\vec{k}) =\frac{i}{\omega} \Big[ \langle J^i J^j\rangle_R(\omega,\vec{k})-\langle J^i J^j\rangle_R(0,\vec{k}) \Big]\;.
\end{equation}
Two steps enter the longitudinal response part of this expression. Kinematically, the factor $1/\omega$ in~\eqref{eq:: E field change of variables} produces a spurious origin pole in the electric-field kernels, cancelled by the two powers of $i\omega$ in
\begin{equation}\label{eq:: JJ from Seff}
     \langle J^i J^j\rangle_R(p) =\frac{\delta^2 S_\text{eff}}{\delta A_i^r(p) \delta A_j^a(-p)} = (i\omega)^2G^R_{ij}(p)
\end{equation}
together with the prefactor $i/\omega$ in~\eqref{eq:: conductivity definition}, leaving a regular conductivity at fixed $k$.

Dynamically, $\psi_l$ remains in~\eqref{eq:: MC effective action}, so $G^R_{||,||}$ is irreducible with respect to  diffusion and defines
\begin{equation}\label{eq:: irreducible conductivity}
    \sigma_{\rm irr}(\omega,\vec k) = \frac{\sigma}{1-i\omega\tau} + \frac{\sigma^2 \tau (2-i\omega\tau)k^2}{\chi (1-i\omega\tau)^2} + O(k^4)\;.
\end{equation}
Here $\sigma_{\rm irr}$ is analytic for $|\omega|<1/\tau$ and reduces at $k=0$ to the single mode memory kernel $B(\omega)=\sigma/(1-i\omega\tau)$. Eliminating $\psi_l$ by a second Schur complement gives the physical correlator,
\begin{equation}\label{eq:: JJ Schur}
    \langle J^{||} J^{||}\rangle_R(p) = (i\omega)^2\Big[\,G^R_{||,||}(p) - G^R_{||l}(p)\big(G^R_{ll}(p)\big)^{-1}G^R_{l||}(p)\Big]\;,
\end{equation}
The subtraction is $O(k^2/\omega)$ and enforces charge conservation at finite momentum. No analogous term appears in the transverse channel.

Carrying out both steps, the conductivity is obtained in closed form in either channel,
\begin{align}
    \sigma^{ij}_\perp (\omega,\vec{k}) &= \frac{\sigma}{1-i\omega\tau} \delta^{ij}\;, \\
    \sigma^{||,||} (\omega,\vec{k}) &= \frac{\sigma}{1-i\omega\tau+\dfrac{ik^2D}{\omega}}
    \;=\; \frac{i\sigma\omega}{\tau\,(\omega-\omega_+)(\omega-\omega_-)}\;, \label{eq:: exact longitudinal conductivity}
\end{align}
with $\omega_\pm$ the Maxwell-Cattaneo frequencies~\eqref{eq:: MC poles appendix} and $D=\sigma/\chi$. The same expression follows immediately from the deterministic equations of motion~\eqref{eq:: MC deterministic eoms} upon eliminating $n$ through the continuity equation, which provides an independent check of the effective action.

Equation~\eqref{eq:: exact longitudinal conductivity} can be rewritten in a form that makes the role of the two steps transparent,
\begin{equation}\label{eq:: sigma from B}
    \sigma^{||,||}(\omega,\vec k) = \frac{\chi\,\omega\,B(\omega)}{\chi\,\omega + ik^2B(\omega)}\;, \qquad B(\omega)=\frac{\sigma}{1-i\omega\tau}\;,
\end{equation}
which is the single-relaxation specialization of~\eqref{Eq:Responses}. The memory kernel is analytic for $|\omega|<1/\tau$ uniformly in $k$, while the full longitudinal response also contains the diffusive denominator. Expanding at small momentum,
\begin{equation}
    \sigma^{||,||} (\omega,\vec{k}) = \frac{\sigma}{1-i\omega\tau} - \frac{i\sigma^2 k^2}{\chi\,\omega\,(1-i\omega\tau)^2} + O(k^4)\;,
\end{equation}
the Drude form  is recovered only for $|\omega|\gg Dk^2$, so the $k^2$ and $\omega$ expansions do not commute in the diffusive window, as in the continuum analysis of Section~\ref{Sec:: Log branch cut}.

\subsection{Perturbative example: $N=2$}
For $N=2$ we illustrate the local reduction perturbatively through $O(k^2)$. Let us start from the action~\eqref{eq:: MC N action} where we choose:
\begin{equation}
    X=\chi_v \begin{bmatrix}
        x_1 & x_m  \\
        x_m & x_2
    \end{bmatrix} \;, \qquad \Sigma =\chi_v \begin{bmatrix}
        \tau & \eta \\
        \eta & \lambda \tau
    \end{bmatrix}\;.
\end{equation}
Positivity of $\Sigma$ imposes
\begin{equation} 
    \lambda\geq0 \;, \qquad \lambda \tau^2 \geq \eta^2\;.
\end{equation}
Then, we need to solve equation~\eqref{Eq:Diffusive mode from matrix determinant lemma}. Perturbatively, up to second order in the momenta, the solutions are:
\begin{subequations}
    \begin{align}
        &\omega_D(k) = -iDk^2 + O(k^4)\\
        &\omega_\pm(k) = \omega_\pm^{(0)} + \omega_\pm^{(2)} k^2 + O(k^4)\;,
    \end{align}
\end{subequations}
where 
\begin{subequations} \label{eq:: Explicit 2 poles values}
    \begin{align}
        D &= \frac{1}{\chi} \vec{\alpha}^t X^{-1} \vec{\gamma}_+\\
        \omega_\pm^{(0)} &=\frac{-i}{2(\lambda \tau^2-\eta^2)} \Bigg[ \tau(\lambda x_1 + x_2) - 2\eta x_m \pm \sqrt{\Delta} \Bigg] \\
        \Delta &=\Big(\tau(\lambda x_1 + x_2)-2\eta x_m\Big)^2 - 4(x_1x_2-x_m^2)\Big(\lambda \tau^2-\eta^2\Big)\;.
    \end{align}
\end{subequations}
We recognize from~\eqref{eq:: MC N constraints from frame choice} the standard result $D=\sigma/\chi$. For the allowed parameters the $k=0$ poles lie on the imaginary axis, as explained in Section~\ref{Sec:: Reality of the branch cuts}, and remain there at sufficiently small momentum until the first collision.

A hierarchy of scales arises whenever one of the two non-hydrodynamic frequencies becomes parametrically smaller than the other. By looking at~\eqref{eq:: Explicit 2 poles values}, we notice that a simple way to realize this regime is by approaching the limit in which $\det [X]$ or $\det [\Sigma]$ becomes small. Under this limit, we can perform a Gaussian integration analogously to what we did for the standard MC case. We perturbatively compute the eigenvectors of
\begin{equation}
    \mathsf{W} = \begin{bmatrix}
        0 & \frac{ik_i}{\chi}\vec{\alpha}^t \\
        \Sigma^{-1}\vec{\gamma}_+ k^i & -i \Sigma^{-1}X
    \end{bmatrix}
\end{equation}
and perform the change of variables
\begin{equation}
    \begin{pmatrix}
        B_\mu \\
        v_{1,\mu} \\
        v_{2,\mu}
    \end{pmatrix} = P(k) \begin{pmatrix}
        \psi_{D,\mu} \\
        \psi_{-,\mu} \\
        \psi_{+,\mu}
    \end{pmatrix}\;.
\end{equation}
We compute the equation of motion by varying $\psi_{+,\mu}^r$ and find an expression for
\begin{equation}
    \psi_{+,\mu}^a = \psi_{+,\mu}^a(\psi_{D,\mu}^a,\psi_{-,\mu}^a)\;.
\end{equation}
It is straightforward to verify that, after the substitution, the terms involving $\psi^r_{+\mu}$ cancel, and we are left with the effective action:
\begin{equation}
    S_{MC}[\psi_D,\psi_-] = \pint \Bigg\{ -\vec{\Phi}_{a,\mu}^t \mathbb{X}_\text{eff}^{\mu\nu} \vec{\Phi}_{r,\nu} + iT\Delta^{\mu\nu}\vec{\Phi}_{a,\mu}^t\mathbb{S}_\text{eff}\tilde{\vec{\Phi}}_{a,\nu}\Bigg\} \;,
\end{equation}
where $\vec{\Phi}_\mu = \begin{pmatrix}\psi_{D,\mu} & \psi_{-,\mu}\end{pmatrix}^t$, while $\mathbb{X}_\text{eff}^{\mu\nu}$ and $\mathbb{S}_\text{eff}$ are the Schur complements of the $(D,-)$ sector:
\begin{equation}
    \mathbb{S}_\text{eff} = \mathbb{S}_{D-} -\mathbb{S}_{D-,+} \mathbb{S}_+^{-1} \mathbb{S}_{+,D-} \;, \qquad \mathbb{S} = \begin{bmatrix}
        \mathbb{S}_{D-} &\mathbb{S}_{D-,+} \\
        \mathbb{S}_{+,D-} &\mathbb{S}_+
    \end{bmatrix} \;.
\end{equation}

The effective theory contains a diffusive mode and a light non-hydrodynamic mode; thus, the action can be rewritten as the standard MC action~\eqref{eq:: Action for standard MC} with $\sigma_\text{eff}$ and $\tau_\text{eff}$.

\bibliographystyle{JHEP}
\bibliography{Bibliography}

\end{document}